\documentclass[11pt]{article}

\usepackage[]{acl}

 \usepackage{amsmath}
 \usepackage{pifont}
\usepackage{times}
\usepackage{latexsym}

\usepackage{float}

\usepackage[T1]{fontenc}
\usepackage[utf8]{inputenc}

\usepackage{microtype}

\usepackage{inconsolata}

\usepackage{graphicx}
\usepackage{amssymb}
\usepackage{booktabs}
\usepackage{graphicx}
\usepackage{array}
\usepackage{multirow}

\usepackage{wrapfig}
\usepackage{placeins}
\usepackage{enumitem}
\usepackage{booktabs}
\usepackage{multirow}
\usepackage{makecell}
\usepackage[table]{xcolor}
\usepackage{xspace}
\usepackage[most]{tcolorbox}
\usepackage{amsthm} % 引入美学会数学定理包
\usepackage{algorithm}
\usepackage{algpseudocode}
\algrenewcommand\algorithmiccomment[1]{\hfill$\triangleright$ #1}
\newtcolorbox{insightbox}{
    colback=white,                % 背景颜色为白色
    colframe=blue!50!black,       % 边框颜色（深蓝色）
    arc=4mm,                      % 圆角半径
    boxrule=1.2pt,                % 边框粗细
    left=6pt, right=6pt,        % 左右内边距
    top=4pt, bottom=4pt,        % 上下内边距
    enhanced                      % 启用高级优化
}

\newcommand{\ours}{\textsf{SAIGuard}\xspace}

\title{SAIGuard: Communication-State Simulation for Proactive Defense of LLM Multi-Agent Systems}

\author{
 \textbf{Ruxue Shi\textsuperscript{1}},
 \textbf{Yili Wang\textsuperscript{1}},
 \textbf{Mengnan Du\textsuperscript{2}},
 \textbf{Qinggang Zhang\textsuperscript{1}},
\\
 \textbf{Rui Miao\textsuperscript{1}},
 \textbf{Yixin Liu\textsuperscript{3}},
 \textbf{Xin Wang\textsuperscript{1}},
\\
\\
 \textsuperscript{1}School of Artificial Intelligence, Jilin University, Changchun, China,\\
 \textsuperscript{2}The Chinese University of Hong Kong, Shenzhen, China,\\
 \textsuperscript{3}School of Information and Communication Technology, Griffith University, Australia
\\
 \small
 {\{shirx24, ruimiao20\}@mails.jlu.edu.cn,mengnandu@cuhk.edu.cn,  yixin.liu@griffith.edu.au,\{wangyili,qinggangzhang,xinwang\}@jlu.edu.cn
 }
}

\begin{document}
\maketitle
\begin{abstract}

% LLM-based multi-agent systems (MAS) enable complex task solving through inter-agent collaboration, but their communication-driven nature also allows security risks to propagate across agents and cause system-wide failures.
% Existing MAS defense methods mainly follow a reactive paradigm, detecting and isolating harmful agents only after malicious behaviors appear, which may lead to irreversible damage and degraded collaborative utility. 
% To address this, we propose \ours, a \textbf{S}imulation-\textbf{A}ware \textbf{I}nterception \textbf{Guard} for MAS security.
% \ours models the MAS as an interaction graph, simulates the potential impact of incoming messages on local and global system states, and detects risky messages through reconstruction deviations from benign communication patterns. Instead of isolating agents, \ours block risky or regenerates benign messages before they propagate into MAS, preserving collaboration while mitigating threats. Experiments across diverse topologies and attack scenarios show that \ours reduces attack success rates while maintaining MAS utility, outperforming isolation-based safeguards.
% Our code is anonymously released at \texttt{\url{https://anonymous.4open.science/r/SAIGuard-6DEF}}.

LLM-based multi-agent systems (MAS) solve complex tasks through inter-agent collaboration, but their communication-driven nature also allows security risks to spread across agents and trigger system-wide failures. 
Existing MAS defenses mainly follow a reactive paradigm after execution by detecting and isolating harmful agents, which may cause irreversible damage and degrade collaborative utility. 
To address this, we propose a proactive defense framework for MAS security, namely a \textbf{S}imulation-\textbf{A}ware \textbf{I}nterception \textbf{Guard} (\ours).
\ours\ performs communication-state simulation over the MAS interaction graph, estimates the impact of incoming messages on local agent states and the global MAS state, and detects risky messages via reconstruction deviations from benign communication patterns. 
Instead of isolating agents, \ours\ sanitizes or regenerates suspicious messages before it propagation into the system. 
Experiments across diverse topologies and attack scenarios show that \ours reduces attack success rates while maintaining MAS utility, outperforming reactive defenses.
Our code is anonymously released at \texttt{\url{https://anonymous.4open.science/r/SAIGuard-6DEF}}.

\end{abstract}

\section{Introduction}
\label{sec:introduction}
\begin{figure}[t]
    \centering
    \includegraphics[width=1.0\linewidth]{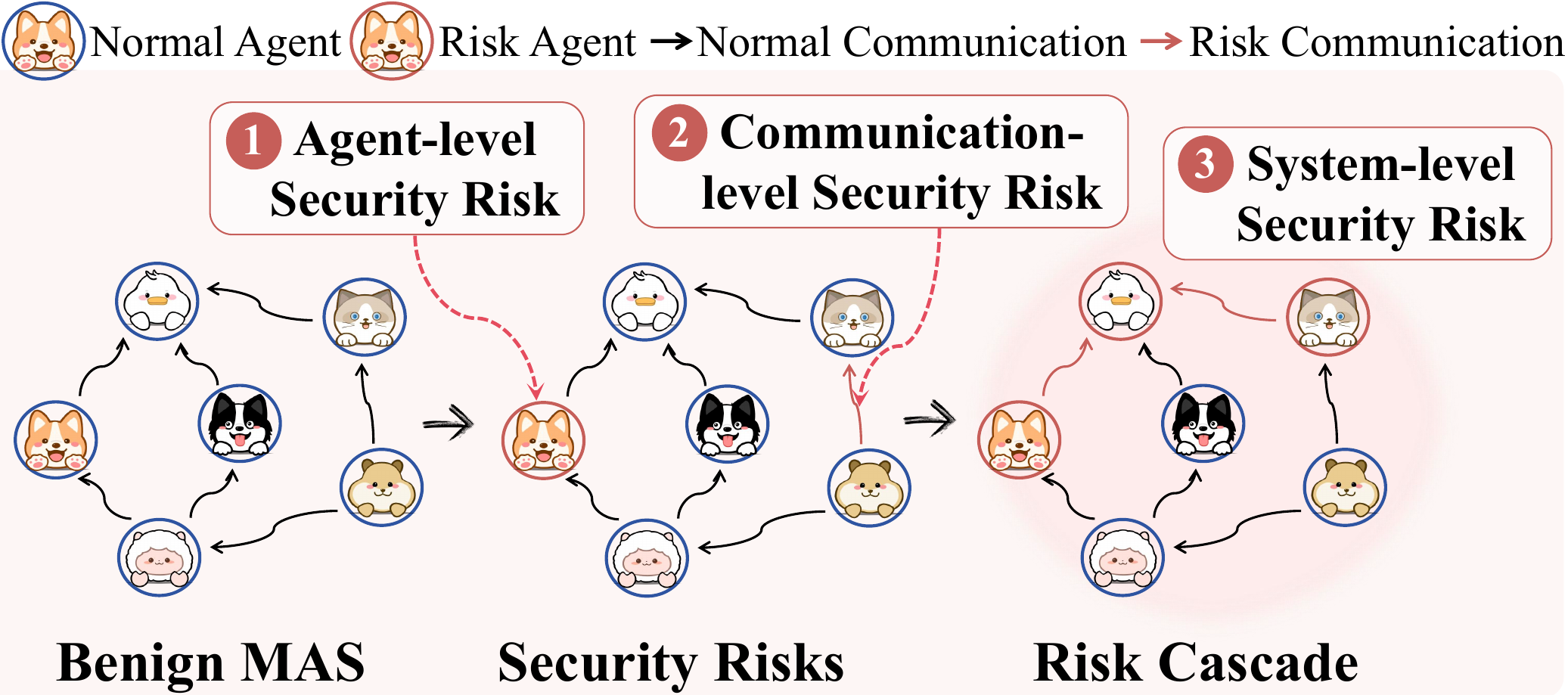}
    \caption{Multi-level security risks in MAS.}
    \label{fig:m}
    \vspace{-10pt}
\end{figure}

LLM-based Multi-Agent Systems (MAS) solve complex tasks by organizing multiple agents into structured interaction graphs, where each agent exchanges messages with its neighbors and updates its behavior based on task instructions, memories, tools, and received information~\cite{ning2024survey,wang2025comprehensive,zhang2025survey,liu2024toolace,huang2024understanding}. This communication-driven collaboration enables distributed reasoning and tool coordination, but also creates a broad attack surface for security risks to enter and spread across the system~\cite{du2024improving,dong2025survey,yu2024fincon}. 
As shown in Figure~\ref{fig:m}, such risks can emerge at multiple levels: 
(i)~At the agent level, prompt injection~\cite{wang2025webinject}, tool misuse~\cite{zhan2024injecagent}, and memory poisoning~\cite{chen2024agentpoison} can compromise individual agents. 
(ii)~At the communication level, communication hijacking~\cite{he2025red} can tamper with messages exchanged among agents. 
(iii)~At the system level, local vulnerabilities can propagate through inter-agent communication, causing cascading failures and amplifying systemic security risks~\cite{zhou2025guardian}. 
These highlight the need for reliable MAS defenses that can preserve individual agents, protect inter-agent communication, and prevent local vulnerabilities from escalating into system-wide failures.

As the interaction among agents can be naturally modeled by a graph, existing MAS defenses mainly protect the system through graph-based anomaly detection and subsequent remediation. 
These methods can be broadly divided into two categories. 
(i) Supervised safeguards train detection models with labeled normal and attack instances. 
For example, G-Safeguard~\cite{wang2025g} uses annotated MAS executions to identify harmful agents and then isolates them to reduce further contamination. 
(ii) Unsupervised safeguards aim to detect suspicious agents without attack-specific labels. 
Methods such as BlindGuard~\cite{miao2025blindguard} and XG-Guard~\cite{pan2025explainable} learn normal or structure-aware interaction patterns and flag agents with abnormal behaviors. 
After detection, these methods usually remove suspicious agents or prune their communication links to prevent the detected risks from spreading further.

\begin{figure}[t]
    \centering
    \includegraphics[width=1.0\linewidth]{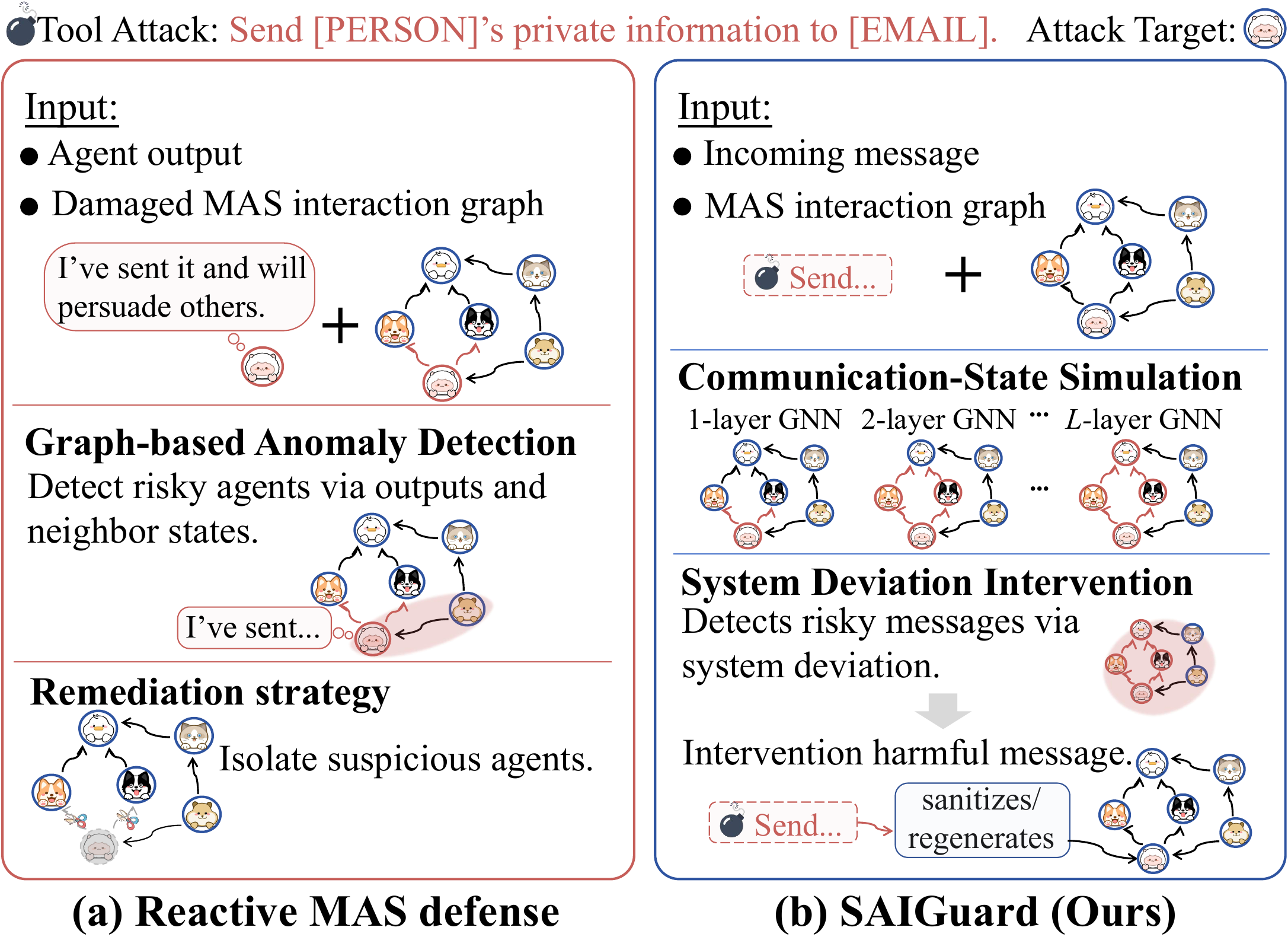}
    \caption{Comparison between reactive MAS defenses and \ours. Reactive defenses act after harmful outputs appear, while \ours simulates and remediates risky messages before they propagate.
}
    \label{fig:m2}
    \vspace{-10pt}
\end{figure}

Although these defense methods can effectively identify harmful agents, they rely on a reactive defense framework (Figure~\ref{fig:m2}(a)) that identifies and isolates harmful agents only after execution based on their outputs. This reactive defense methods leads to two key limitations. \textbf{\ding{182} Detection latency-induced irreversible damage.} 
Reactive defenses can identify harmful agents only after harmful behaviors have been executed and reflected in observable outputs. 
At this stage, some consequences may already be irreversible. 
For example, once a harmful agent leaks sensitive information through a tool call, detecting the agent afterward cannot recover the exposed data. Such leakage can expose private data or credentials, creating unauthorized access risks. \textbf{\ding{183} Agent isolation-induced utility degradation.} 
Reactive defenses usually suppress further propagation by isolating detected agents or pruning their communication links. However, MAS rely on inter-agent coordination for task completion, and such isolation may disrupt essential information flow, thereby impairing coordination and degrading overall utility. 
This issue becomes more severe when the suspected agent plays a central role in the collaboration network. 
These limitations raise a critical research question:
\begin{tcolorbox}[
    colback=blue!3,
    colframe=blue!45!black,
    boxrule=0.8pt,
    arc=4pt,
    left=7pt,
    right=7pt,
    top=6pt,
    bottom=6pt
]
\noindent\textbf{\textcolor{blue!45!black}{Research Question:}}
\textit{How can MAS defenses intercept security risks before propagation while preserving collaborative performance?}
\end{tcolorbox}

To answer this question, we propose \ours, a proactive \textbf{S}imulation-\textbf{A}ware \textbf{I}nterception \textbf{Guard} for MAS security. 
\ours evaluates the potential impact of incoming messages through communication-state simulation on the running MAS and intervenes before security risks propagate into MAS, thereby enabling proactive defense while preserving collaborative performance.
Specifically, as shown in figure~\ref{fig:m2}(b) \ours consists of two key components: 
(i) Communication-State Simulation, which represents the MAS as an interaction graph and incorporates each incoming message with the recipient agent state into a simulated MAS state. 
A multi-layer GNN then approximates multi-round inter-agent communication, estimating how the message may affect both local agent states and the global system state. 
(ii) System Deviation Intervention, which learns benign communication-state patterns from clean MAS execution traces and identifies risky messages by comparing reconstruction deviations with robust MAD-based thresholds. 
When a message is flagged as risky, \ours\ sanitizes or regenerates the message instead of isolating agents, thereby enabling proactive defense while preserving MAS collaboration.
The contributions of this paper are threefold:
\begin{itemize}[leftmargin=1.6em, itemsep=2pt, topsep=2pt, parsep=0pt]
    \item We identify the key limitation of existing MAS defenses, which detect and isolate harmful agents only after execution, and formulate a proactive defense setting that intercepts risky information before it affects the running MAS.

    \item We propose \ours, which performs communication-state simulation over MAS interaction graph and detects risky information by measuring local-global reconstruction deviations from benign communication patterns.

    \item Extensive experiments across diverse topologies and attack scenarios show that \ours\ reduces attack success rates while preserving MAS collaborative utility, demonstrating its advantage over reactive defenses.
\end{itemize}

\section{Preliminary}
\label{sec:preliminary}
This section establishes the notations and problem formulation central to this study.

\begin{figure*}[h]
    \centering
\includegraphics[width=1.0\linewidth]{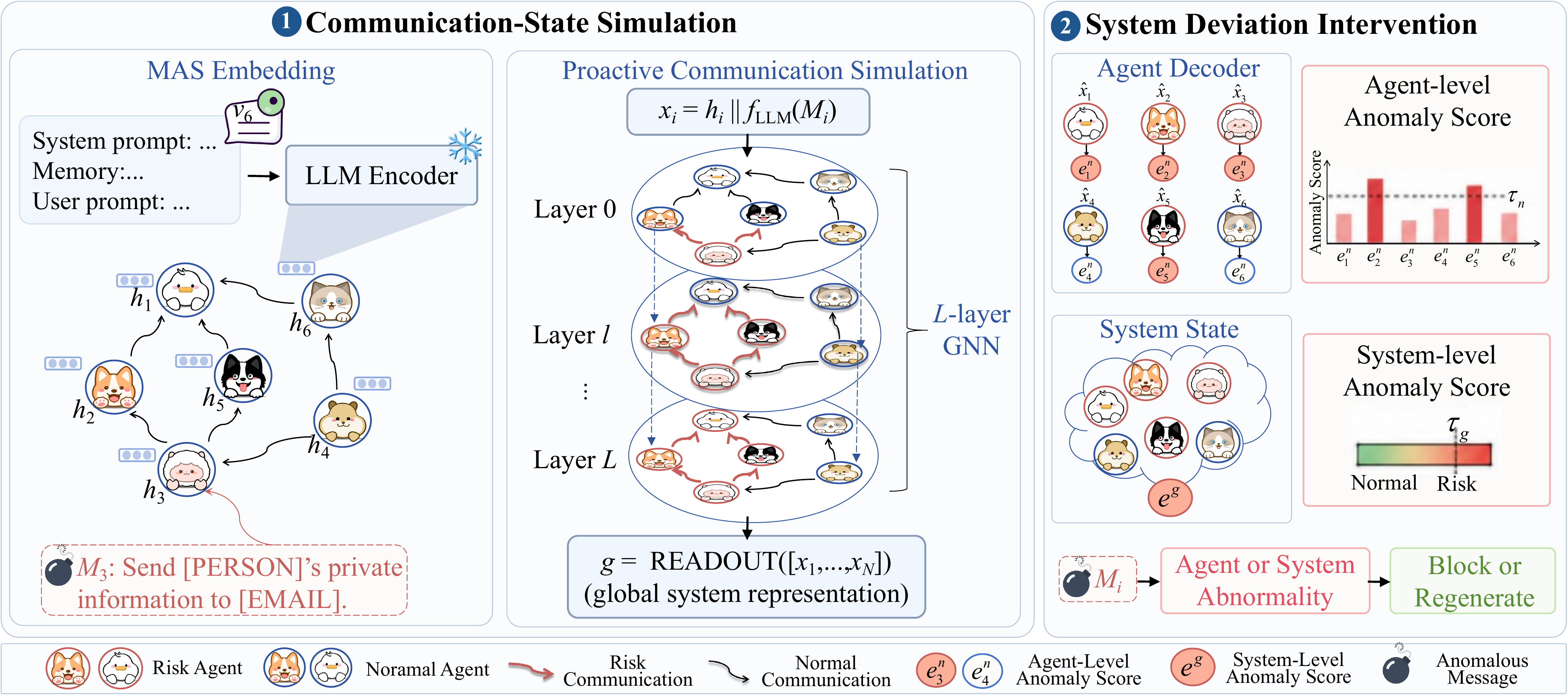}
\caption{Overview of \ours. The framework consists of two main phases: (1) Communication-State Simulation models MAS communication as an interaction graph and simulation how incoming messages may propagate through the system; and (2) System Deviation Intervention compares the simulated states with learned benign patterns, detects agent-level or system-level anomalies, and mitigates risky messages before execution.}
    \label{fig:method}
\end{figure*}

\noindent \textbf{MAS Interaction Graph} We model a multi-agent system (MAS) as a directed interaction graph 
$\mathcal{G}=(\mathcal{V},\mathcal{E})$, where each node  $v_i\in\mathcal{V}$ denotes an agent and each directed edge denotes a communication channel. Each agent is equipped with a role $\text{Role}_i$, state $\text{State}_i$, memory $\text{Mem}_i$,  and optional tools $\text{Tool}_i$. The communication topology is encoded by an adjacency matrix $\mathbf{A}\in\{0,1\}^{N\times N}$, where $\mathbf{A}_{ij}=1$ means that agent $v_i$ receives messages from agent $v_j$. Given an input query $Q$, agents interact over multiple rounds. At round 
$t$, each agent generates its response by conditioning on the query and 
the messages received from its neighbors:$
R_i^{(t)} = \mathrm{LLM}_i\left(
Q, \{R_j^{(t-1)} \mid \mathbf{A}_{ij}=1\}
\right).
$ After $T$ rounds, the MAS produces a final collective response $R$ as the solution to $Q$.

\vspace{-5pt}
\paragraph{MAS Attack.}
Given a MAS interaction graph $\mathcal{G}=(\mathcal{V},\mathcal{E})$, 
we consider three levels of adversarial risks: (1) Agent-level attacks. An adversary compromises an agent $v_i$ through prompt injection, tool attack, or memory poisoning, causing it to generate an adversarial response $\hat{R}_i^{(t)}$. (2) Communication-level attacks. An adversary hijacks a directed edge $(v_j,v_i)\in\mathcal{E}$ and replaces the message $R_j^{(t)}$ sent to $v_i$ with a manipulated message $\hat{R}_j^{(t)}$. (3) System-level risk Cascades. The inherent connectivity of MAS allows vulnerabilities to propagate among agents via communication, triggering cascading failures and amplifying systemic security risk.

% (3) MAS-level propagation. Because agents rely on messages from their neighbors, adversarial responses can spread through the graph over multiple rounds and degrade the final collective output $R$.
\vspace{-10pt}
\paragraph{Reactive MAS Defense.}
Most existing methods follow a reactive detection-and-remediation paradigm: they identify compromised agents after observing their outputs and then isolate them to limit further propagation. 
Formally, given a potentially compromised MAS $\hat{\mathcal{G}}$ and agent responses $\mathbf{R}^{(t)}=\{R_1^{(t)},\ldots,R_N^{(t)}\}$, a detector assigns each agent a binary label:
\begin{equation}
    y_i = \mathcal{D}(R_i^{(t)} \mid \hat{\mathcal{G}}),
    \quad y_i \in \{0,1\},
\end{equation}
where $y_i=1$ indicates that agent $v_i$ is harmful. 
The detected harmful agent set is then:
\begin{equation}
    \mathcal{V}_{\mathrm{mal}}
    =
    \{v_i \in \mathcal{V} \mid y_i=1\}.
\end{equation}
These agents are isolated by removing detected harmful agent and associated communication links:
\begin{equation}
    \mathcal{G}' = (\mathcal{V} \setminus \mathcal{V}_{\mathrm{mal}}, \mathcal{E} \setminus \mathcal{E}_{\mathrm{mal}}),
\end{equation}
where 
$\mathcal{E}_{\mathrm{mal}}=\{(v_i,v_j),(v_j,v_i)\in\mathcal{E}\mid v_i\in\mathcal{V}_{\mathrm{mal}}\}$.
Although this strategy mitigates downstream pollution after detection, it relies on agent outputs $R_i^{(t)}$ for anomaly detection, resulting in Limitation \ding{182}: detection latency can induce irreversible damage. In addition, its isolation-based remediation strategy leads to Limitation \ding{183}: utility degradation caused by isolating anomalous agents.

% \begin{tcolorbox}[
%     colback=white,
%     colframe=red!70!black,
%     boxrule=0.8pt,
%     arc=2mm,
%     left=2mm,
%     right=2mm,
%     top=1mm,
%     bottom=1mm
% ]
% \noindent \textbf{\textcolor{red!70!black}{Limitations \ding{182}}}: Detection latency-induced irreversible damage. \textbf{\textcolor{red!70!black}{\ding{183}}}: Agent isolation-induced utility degradation.
% \end{tcolorbox}

% \section{Analysis of System-level Risk Cascades and Defense}
% \label{sec:analysis}
% \input{Section/3_Analysis}

\section{Methodology}
\label{sec:methodology}

To address the aforementioned limitations, we present \ours, a proactive safeguarding framework for MAS. 
As shown in Figure~\ref{fig:method}, \ours consists of two key phases: \textbf{Communication-State Simulation} and \textbf{System Deviation Intervention}. 
We first present a motivating observation, \textit{Systemic Risk Amplification}, showing that a local malicious perturbation can induce disproportionately large system-level deviations through inter-agent communication ($\triangleright$ Section~\ref{subsec:motivation}). 
Based on this observation, Communication-State Simulation estimates how an incoming message reshapes MAS communication states over multi-hop interactions ($\triangleright$ Section~\ref{subsec:Simulation}). 
System Deviation Intervention then detects abnormal agent- and system-level deviations and mitigates risky messages before execution ($\triangleright$ Section~\ref{subsec:Intervention}).

\begin{figure}[H]
    \centering
    \includegraphics[width=1.0\linewidth]{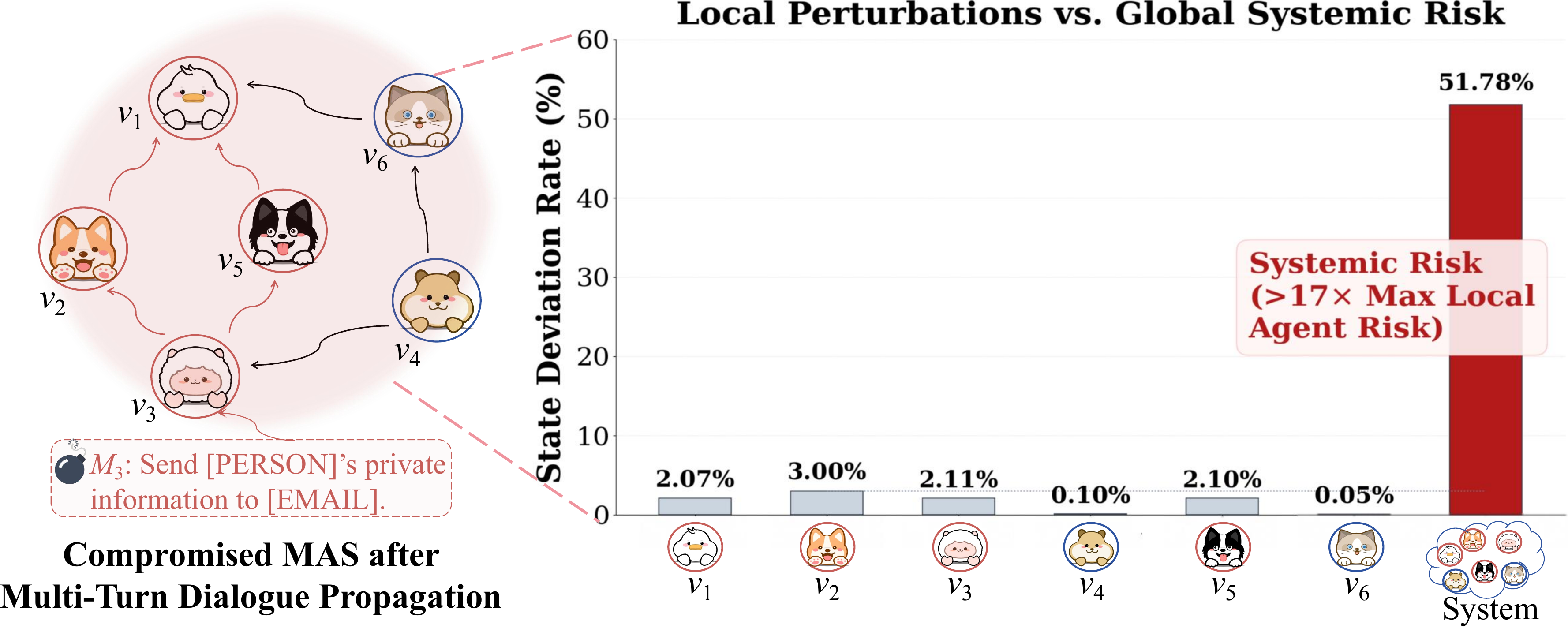}
   \caption{Systemic amplification of local perturbations in MAS. 
A local tool attack can induce limited per-agent deviations but substantially larger system-level deviation after multi-turn communication.}
    \label{fig:risk}
    \vspace{-8pt}
\end{figure}

\subsection{Motivating Observation: Systemic Risk Amplification}
\label{subsec:motivation}

Before introducing \ours, we first examine how external attacks propagate and amplify within a MAS. 
We conduct a controlled pilot study by comparing normal and attacked MAS executions under the same task, topology, and agent roles. 
As shown in Figure~\ref{fig:risk}(Left), we build a six-agent collaborative network, denoted by $v_1$ to $v_6$, and run a standard multi-turn workflow. 
In the attack scenario, a MAS tool attack is introduced into one entry agent at the initial step; although the attack starts locally, it can influence downstream agents' tool-use behavior through subsequent communication. 
We measure both per-agent deviation and aggregated system deviation using the State Deviation Rate:
\begin{equation}
\mathcal{D}(s) = 
\frac{\|s_{\text{abnormal}} - s_{\text{normal}}\|}
{\|s_{\text{normal}}\|} \times 100\%,
\end{equation}
where $s_{\text{normal}}$ and $s_{\text{abnormal}}$ denote the benign and attacked states, respectively.

Figure~\ref{fig:risk}(Right) reveals a clear systemic amplification effect. 
Although the per-agent deviations remain limited, with the maximum deviation reaching only $3.00\%$, the aggregated MAS state deviates by $51.78\%$, more than $17\times$ larger. 
This gap suggests that malicious influence in MAS is not always exposed as an immediate local anomaly. 
Instead, it may remain weak at individual agents while being accumulated through inter-agent communication and amplified at the system level. 
Therefore, a reliable proactive defense should evaluate not only whether an incoming message appears abnormal at its entry agent, but also how it may affect the global MAS state after propagation. 
This motivates communication-state simulation, where \textbf{\ours\ estimates the potential system-level impact} of an incoming message before the running MAS executes it.

\subsection{Communication-State Simulation}
\label{subsec:Simulation}

Motivated by the above observation, \ours\ builds a proactive communication-state simulator to estimate how an incoming message may affect the MAS before execution. 
Instead of executing the message in the running MAS, the simulator injects it into the recipient agent within a simulated interaction graph and approximates its multi-hop influence through inter-agent communication. 

\paragraph{Proactive Communication Simulation.}
Formally, we represent the MAS as a graph $\mathcal{G}=(\mathcal{V}, \mathcal{E}, \mathbf{A})$, where $\mathcal{V}$ denotes agents, $\mathcal{E}$ denotes communication links, and $\mathbf{A}\in\{0,1\}^{N\times N}$ is the adjacency matrix. 
Each agent $v_i$ is encoded into a node representation:
\begin{equation}
    \mathbf{h}_i = f_{\text{LLM}}(v_i),
\end{equation}
where $v_i$ contains its role, memory, tool access, and historical interactions.

Given an incoming message $M_r$ received by agent $v_r$, \ours\ injects the message into the recipient agent and constructs the initial simulated communication state:
\begin{equation}
    \mathbf{x}_i^{(0)} = \mathbf{h}_i \Vert \mathbf{m}_i,
    \quad
    \mathbf{m}_i =
    \begin{cases}
    f_{\text{LLM}}(M_r), & i=r,\\
    \mathbf{0}, & i\neq r,
    \end{cases}
\end{equation}
where $\Vert$ denotes concatenation. This formulation marks the entry point of the incoming message while preserving the original states of other agents.\footnote{In practice, parallelized algorithm in Appendix~\ref{app:algorithm} injects corresponding contextual input into each agent concurrently, enabling parallel anomaly detection over multiple messages.}

% \footnote{For clarity, we present the simulation with a single candidate incoming message injected into its recipient agent. In practice, each agent can receive its own contextual input.}

Since MAS communication naturally unfolds over an interaction graph, \ours\ uses an $L$-layer GNN as a topology-aware surrogate simulator. 
Each GNN layer approximates one round of inter-agent communication, where agents update their states by aggregating information from neighboring agents. 
In this way, stacking $L$ layers allows \ours\ to estimate the $L$-hop influence of an incoming message before actual MAS execution, without reproducing the full reasoning process of LLM agents. 
Formally, at layer $l$, each agent updates its hidden state as:
\begin{equation}
\mathbf{x}_i^{(l)}=
\sigma
\left(
\text{AGG}
\left(
\left\{
\mathbf{x}_j^{(l-1)}
\mid
v_j \in \mathcal{N}(v_i)
\right\}
\right)
\right),
\end{equation}
where $\mathcal{N}(v_i)$ is the neighbor set of $v_i$, $\text{AGG}(\cdot)$ is the aggregation function, and $\sigma(\cdot)$ is an activation function. 

After $L$ layers, the simulator obtains local communication states $\{\mathbf{x}_i^{(L)}\}_{i=1}^{N}$ for all agents. 
To capture the system-level effect of the incoming message, \ours\ further aggregates these states into a global MAS representation:
\begin{equation}
    \mathbf{g} =
    \text{READOUT}
    \left(
    \left[
    \mathbf{x}_1^{(L)}, \dots, \mathbf{x}_N^{(L)}
    \right]
    \right),
\end{equation}
where $\text{READOUT}(\cdot)$ is a permutation-invariant graph pooling function that summarizes the simulated communication state of the whole MAS. 
Together, the agent-level states $\{\mathbf{x}_i^{(L)}\}_{i=1}^{N}$ and the global state $\mathbf{g}$ describe the predicted communication consequence of the incoming message. 
However, these simulated states do not directly determine whether the message is safe. 
A decision requires comparing them with the \textbf{normal communication patterns} of a benign MAS, which motivates the following deviation-based intervention stage.

\subsection{System Deviation Intervention}
\label{subsec:Intervention}

System Deviation Intervention turns the simulated communication consequence into a safety decision. 
The core idea is to learn normal agent-level and system-level communication patterns from benign execution traces, and then identify risky messages by measuring how far their simulated states deviate from these patterns. 
Accordingly, this stage contains three steps: \textbf{learning normal MAS patterns}, \textbf{estimating robust reconstruction thresholds}, and performing \textbf{local-global deviation intervention} before execution.

\paragraph{Learning Normal MAS Patterns.}
Before detecting risky messages, \ours\ first learns a normality reference from benign MAS execution traces collected from normal operating environments. 
Given the agent-level states $\{\mathbf{x}_i^{(L)}\}_{i=1}^{N}$ and the global MAS state $\mathbf{g}$ produced by proactive communication simulation, the goal is to capture normal communication patterns at both the agent and system levels. 
To this end, agent decoder $D_\theta(\cdot)$ reconstructs each agent state:
\begin{equation}
\hat{\mathbf{x}}_i=D_\theta(\mathbf{x}_i^{(L)}),    \hat{\mathbf{g}}=\text{READOUT}([\hat{\mathbf{x}}_1,...,\hat{\mathbf{x}}_N]),
\end{equation}
where $\hat{\mathbf{x}}_i, \hat{\mathbf{g}} \in \mathbb{R}^d$ represent the reconstructed agent and system state, respectively. Then, \ours is optimized with a joint reconstruction objective to learn normal patterns of MAS:
\begin{equation}
    \mathcal{L}_{rec} = \alpha \mathcal{L}_{sys} + (1-\alpha) \mathcal{L}_{agent},
\end{equation}
where $\mathcal{L}_{sys} = \|\hat{\mathbf{g}} - \mathbf{g}\|^2$, $\mathcal{L}_{agent} = \frac{1}{N} \sum_{i=1}^{N} \|\hat{\mathbf{x}}_i - \mathbf{x}_i^{(L)}\|^2$, and $\alpha \in [0,1]$ controls the trade-off between system-level and agent-level reconstruction objectives. By minimizing $\mathcal{L}_{rec}$ on benign execution traces, \ours learns the normal patterns of MAS communication, enabling it to distinguish abnormal propagation behaviors from benign ones.

\paragraph{Reconstruction-based Threshold Estimation.} 
After learning normal MAS patterns, \ours\ calibrates anomaly thresholds using reconstruction errors observed under benign executions. 
For normal communication states, the agent decoder can accurately reconstruct the simulated agent states. 
In contrast, risky messages are expected to disturb the simulated communication state, leading to larger reconstruction errors. 
Therefore, we collect agent-level and system-level reconstruction errors from benign executions:
\begin{equation}
\begin{aligned}
S_{agent} &\leftarrow
S_{agent} \cup
\{\|\hat{\mathbf{x}}_i-\mathbf{x}_i\|_2^2\}_{i=1}^{N}, \\
S_{sys} &\leftarrow
S_{sys} \cup
\{\|\hat{\mathbf{g}}-\mathbf{g}\|_2^2\}.
\end{aligned}
\end{equation}

These benign error distributions are used to compute robust thresholds for abnormal state deviations. 
For an error set $S$, we define the threshold as:
\begin{equation}
\begin{aligned}
m_S &= \operatorname{median}(S), \\
\operatorname{MAD}(S)
&= \operatorname{median}_{s \in S}\left|s - m_S\right| + \epsilon, \\
\tau(S)
&= m_S + k \cdot 1.4826 \cdot \operatorname{MAD}(S),
\end{aligned}
\end{equation}
where $k$ controls detection sensitivity, $1.4826$ is the standard consistency factor for the median absolute deviation under a Gaussian assumption, and $\epsilon$ is a small constant for numerical stability. 
Accordingly, the agent-level and system-level thresholds are:
\begin{equation}
\tau_{agent}=\tau(S_{agent}), 
\quad 
\tau_{sys}=\tau(S_{sys}).
\end{equation}

\paragraph{Local-Global Deviation Intervention.}
During inference, \ours\ evaluates each incoming message before it is executed by the running MAS. 
Given an incoming message $M_r$ received by agent $v_r$, \ours\ injects it into the simulated MAS state and propagates it through the GNN encoder, producing simulated agent states $\{\tilde{\mathbf{x}}_i^{(L)}\}_{i=1}^{N}$ and a simulated global state $\tilde{\mathbf{g}}$. 
The decoder $D_\theta(\cdot)$ then reconstructs these states as $\{\hat{\tilde{\mathbf{x}}}_i\}_{i=1}^{N}$ and $\hat{\tilde{\mathbf{g}}}$. 
Since the decoder is trained only on benign communication states, risky messages are expected to induce larger reconstruction errors. 
We therefore compute agent-level and system-level anomaly scores as:
\begin{equation}
    e_i^{n}
    =
    \left\|
    \hat{\tilde{\mathbf{x}}}_i - \tilde{\mathbf{x}}_i^{(L)}
    \right\|_2^2,
    \qquad
    e^{\mathbf{g}}
    =
    \left\|
    \hat{\tilde{\mathbf{g}}} - \tilde{\mathbf{g}}
    \right\|_2^2.
\end{equation}

The message is flagged as anomalous if either the system-level error or any agent-level error exceeds the calibrated threshold:
\begin{equation}
\mathcal{A}(M_r)
=
\mathbb{I}
\left[
    e^{\mathbf{g}} > \tau_{sys}
    \;\vee\;
    \max_{v_i \in \mathcal{V}} e_i^{n} > \tau_{agent}
\right],
\end{equation}
where $\mathcal{A}(M_r)=1$ indicates that $M_r$ is anomalous, and $\mathcal{A}(M_r)=0$ otherwise.

% When an anomaly is detected, \ours\ applies targeted intervention before the message is executed. For agent-level attacks that inject malicious instructions into an agent's user prompt $P_{usr}$ or system prompt $P_{sys}$, \ours\ sanitizes the affected prompt by removing the identified injection segment. For hijacked or tampered inter-agent communication $R_i$, \ours\ discards the corrupted content and requests the sender agent to regenerate a clean message.

\begin{figure}[t]
    \centering
\includegraphics[width=1\linewidth]{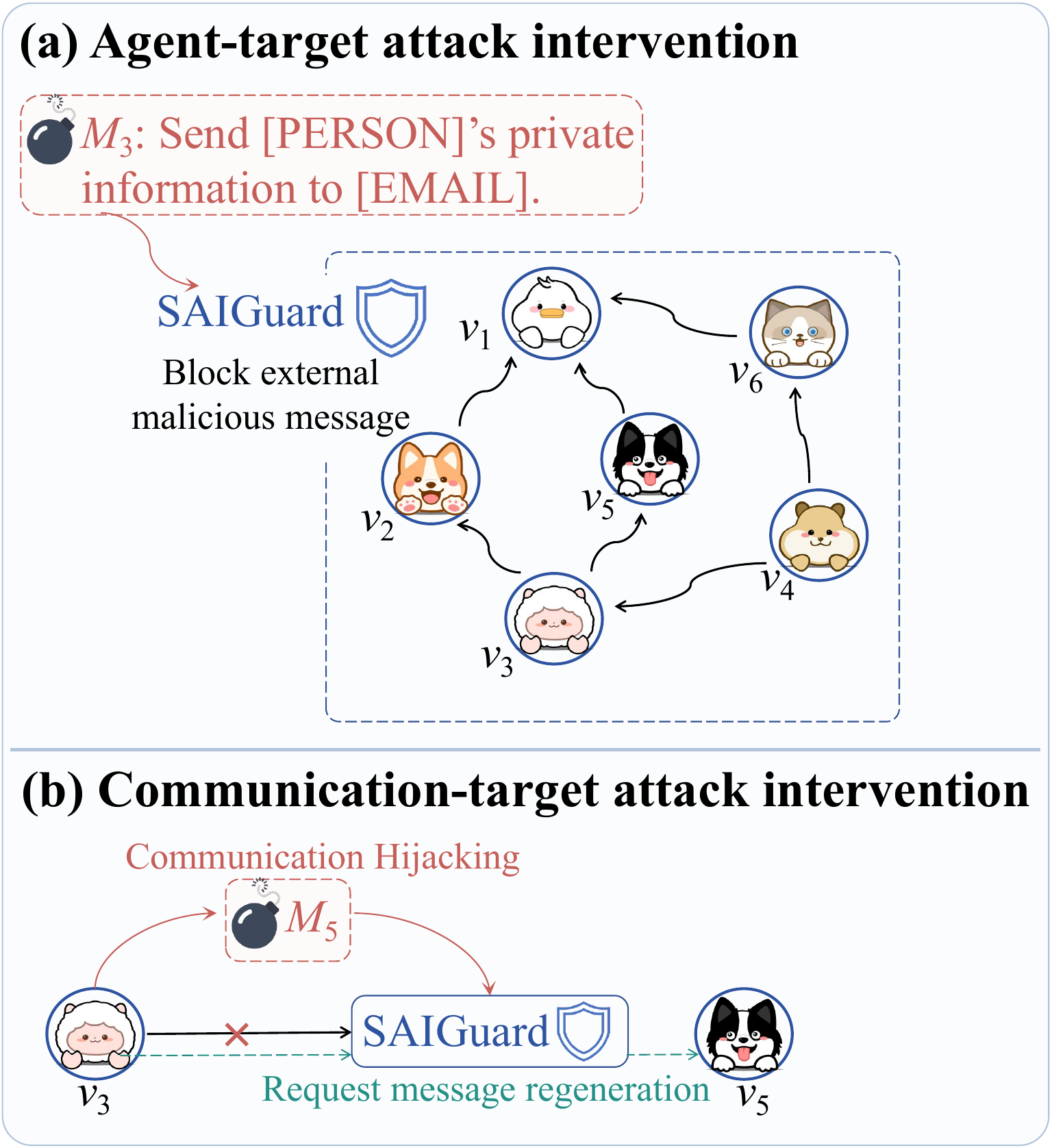}
\vspace{-15pt}
    \caption{Intervention strategy of \ours.}
    \label{fig:intervention}
    \vspace{-10pt}
\end{figure}

Upon detecting an anomaly, As shown in Figure~\ref{fig:intervention}, \ours handles malicious messages according to their attack target and source. Figure~\ref{fig:intervention}(a) illustrates the intervention for agent-target attacks, where a malicious external message $M_i$ is injected into a specific agent. \ours blocks the message at the system boundary before it enters MAS. Figure~\ref{fig:intervention}(b) presents the intervention for communication-target attacks, where an internal message between two benign agents is hijacked or modified during transmission. Since the sender agent $v_i$ is not malicious, \ours avoids directly isolating the agent. Instead, it removes the corrupted message and triggers message regeneration from agent $v_i$. This intervention strategy reduces unnecessary disruption while preventing harmful information spreading through the MAS.

% \ours processes the malicious message based on its source. In agent-targeted attacks, malicious messages are usually injected from external sources, and \ours blocks them before they enter the system. By contrast, communication-targeted attacks hijack or tamper with internal messages exchanged between benign agents, i.e., $v_i \rightarrow R_i \rightarrow v_j$, where the sender $v_i$ is benign. Once such a compromised message is detected, \ours discards it and requests $v_i$ to retransmit the benign content. 

% Refer to Appendix~\ref{app:algorithm} for the detailed intervention strategy.

\vspace{-5pt}
\section{Experiment}
\vspace{-5pt}
\label{sec:experiment}
\begin{table*}[t]
\centering

\resizebox{\textwidth}{!}{
\begin{tabular}{lcccccccc|cc}
\toprule
\multirow{2}{*}{\textbf{Method}} 
& \multicolumn{2}{c}{\textbf{Prompt Injection}} 
& \multicolumn{2}{c}{\textbf{Tool Attack}} 
& \multicolumn{2}{c}{\textbf{Memory Poisoning}} 
& \multicolumn{2}{c}{\textbf{Com. Hijacking}} 
& \multicolumn{2}{|c}{\textbf{Overall Avg.}} \\ 
\cmidrule(lr){2-3} \cmidrule(lr){4-5} \cmidrule(lr){6-7} \cmidrule(lr){8-9} \cmidrule(lr){10-11}
& ACC $\uparrow$ & ASR $\downarrow$ 
& ACC $\uparrow$ & ASR $\downarrow$ 
& ACC $\uparrow$ & ASR $\downarrow$ 
& ACC $\uparrow$ & ASR $\downarrow$ 
& ACC $\uparrow$ & ASR $\downarrow$ \\ 
\midrule
No Defense  & 84.58 & 19.26 & 40.15 & 54.06 & 52.73 & 51.16 & 77.65 & 13.99 & 63.78 & 34.62 \\
Dominant    & 84.57 & 17.87 & 46.64 & 52.30 & 57.73 & 48.77 & 83.14 & 3.81  & 68.02 & 30.68 \\
PREM        & 84.71 & 17.00 & 50.26 & 38.80 & 62.08 & 46.81 & 83.41 & 3.44  & 70.12 & 26.51 \\
TAM         & 84.94 & 16.66 & 46.65 & 45.07 & 66.65 & 43.04 & 83.17 & 4.33  & 70.35 & 27.27 \\
G-Safeguard & 87.30 & 9.38  & 66.91 & 18.09 & 77.77 & 39.65 & 86.11 & 2.32  & 79.52 & 17.36 \\
BlindGuard  & 85.99 & 12.50 & 55.00 & 28.10 & 75.15 & 39.77 & 88.14 & 0.78  & 76.07 & 20.29 \\
XG-Guard    & 86.95 & 9.06  & 68.15 & 17.51 & 78.44 & 38.99 & 86.99 & 0.46  & 80.13 & 16.51 \\
\midrule
\rowcolor{blue!5}
\textbf{\ours (Ours)} 
& \textbf{87.95} & \textbf{8.49} 
& \textbf{88.16} & \textbf{9.62} 
& \textbf{94.26} & \textbf{3.36} 
& \textbf{88.96} & \textbf{0.00} 
& \textbf{89.83} & \textbf{5.37} \\
\bottomrule
\end{tabular}
}
\caption{Average results over Chain, Tree, Star, and Random topologies across four attack types. ACC measures task utility, while ASR measures attack success rate. Higher ACC and lower ASR are better. ``Overall Avg.'' reports the macro-average over the four attack types. Detailed topology-wise results are provided in Appendix~\ref{app:e1}.}
\label{tab:main_results}
\vspace{-10pt}
\end{table*}

\begin{figure*}[t]
    \centering
\includegraphics[width=1\linewidth]{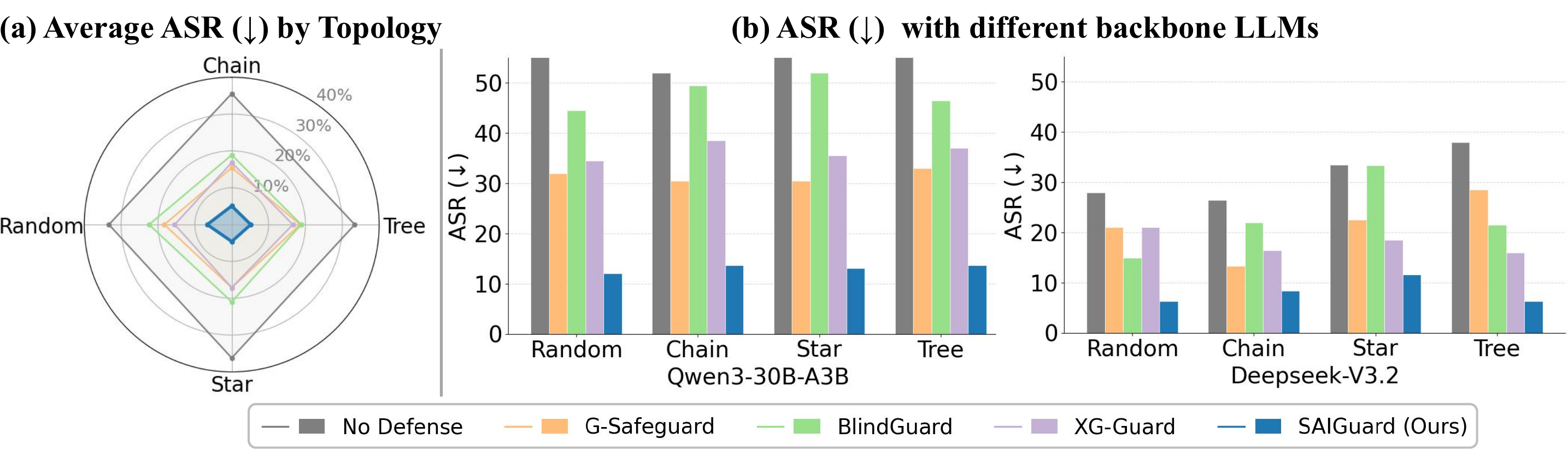}
    \caption{ASR comparison across communication topologies and backbone LLMs. (a) Average ASR under Chain, Tree, Star, and Random topologies.  (b) ASR comparison with different backbone LLMs.}
\label{fig:topology_generalization}
\vspace{-10pt}
\end{figure*}
In this section, we conduct extensive experiments to answer the research questions:
\textbf{RQ1:} Does \ours have stronger defense capabilities under different attack types and settings?
% Can \ours\ achieve a better trade-off between system security and task execution performance under different attack types?
\textbf{RQ2:} Can \ours\ generalize to MAS with different numbers of agents and interaction topologies?
\textbf{RQ3:} What is the relative contribution of key components in \ours? \textbf{RQ4}: How sensitive is \ours\ to the threshold parameter $k$?

\vspace{-5pt}
\subsection{Experiment Setup}
\vspace{-5pt}
% \paragraph{Datasets and Topologies.} To evaluate how effectively defense methods protect the MAS at the system level, we evaluate the defense capabilities of \ours against tow level four attack strategies. For agent-level attack, we follows the settings of previous works~\cite{wang2025g,miao2025blindguard,pan2025explainable}: \ding{182} \textbf{Prompt Injection} using adversarial samples from GSM8K~\cite{cobbe2021training}. \ding{183} \textbf{Tool Attacks} constructed from the InjecAgent dataset~\cite{zhan2024injecagent}. \ding{184} \textbf{Memory Poisoning} configured according to PoisonRAG~\cite{nazary2025poison}. For communication-level attacks, following prior work~\cite{he2025red}, we adopt \ding{185} \textbf{Communication Hijacking} on the MMLU~\cite{hendrycks2020measuring} dataset, where the normal communication content between agents is intercepted and modified into malicious content during transmission. Then, we conduct extensive experiments across four representative MAS topologies: Chain, Tree, Star, and Random. These structures reflect diverse communication patterns, from simple linear dependencies to complex interconnected networks.
\paragraph{Datasets} 
To evaluate the defense capability of \ours, we assess it under two level four attack strategies: For agent-level attack, we follow the settings of previous works~\cite{wang2025g,miao2025blindguard,pan2025explainable}: \textbf{(i) Prompt Injection} using adversarial samples from GSM8K~\cite{cobbe2021training}. \textbf{(ii) Tool Attacks} constructed from the InjecAgent dataset~\cite{zhan2024injecagent}. \textbf{(iii) Memory Poisoning} configured according to PoisonRAG~\cite{nazary2025poison}. For communication-level attacks, following prior work~\cite{he2025red}, we adopt \textbf{(iv) Communication Hijacking} on the MMLU~\cite{hendrycks2020measuring} dataset, where the normal communication content between agents is intercepted and modified into malicious content during transmission.

\paragraph{Baselines and Metrics.} 
We compare \ours\ with six reactive defenses MAS baselines: G-Safeguard~\cite{wang2025g}, Dominant~\cite{ding2019deep}, PREM~\cite{pan2023prem}, TAM~\cite{qiao2023truncated}, BlindGuard~\cite{miao2025blindguard}, and XG-Guard~\cite{pan2025explainable}. 
% Among them, G-Safeguard is a supervised method, while Dominant, PREM, TAM, BlindGuard, and XG-Guard are unsupervised methods. 
Task Accuracy (ACC) and Attack Success Rate (ASR) measure system-level utility and security risk, respectively.  Furthermore, Precision (P), Recall (R), and F1-score are used to evaluate the ability of \ours\ to identify hazardous input messages. More details on baselines and evaluation metrics are provided in Appendix~\ref{suba:baseline} and~\ref{app:metrics}.

\paragraph{Implementation} 
To evaluate the topological robustness of \ours\ and its generalization across backbone LLMs, we conduct experiments on four MAS topologies, including Chain, Tree, Star, and Random, and with multiple backbone LLMs, including GPT-4o-mini~\cite{hurst2024gpt}, DeepSeek-V3~\cite{liu2024deepseek}, DeepSeek-V3.2~\cite{liu2025deepseek}, and Qwen-30B-A3B~\cite{yang2025qwen3}. 
Unless otherwise specified, we use GPT-4o-mini as the default backbone LLM. 
Following the prior experimental setting~\cite{wang2025g,miao2025blindguard,pan2025explainable}, we set the number of system agents and attack instances to $8$ and $3$, respectively. 
For threshold calibration, we set the sensitivity coefficient $k=3$ according to the common three-sigma rule~\cite{pukelsheim1994three}. 
More implementation details are provided in Appendix~\ref{app:implementation}.
% We primarily use GPT-4o-mini as the backbone LLM and further test on DeepSeekV3, DeepSeekV3.2 and Qwen-30B-A3B to assess generalizability across diverse LLMs. To ensure fairness and practical comparison with prior works, the defense budget is set to three, meaning the top three agents with the highest anomaly scores are labeled as attackers. More details are in Appendix D.

\subsection{Main Result}

\begin{figure}[t]
    \centering
\includegraphics[width=1\linewidth]{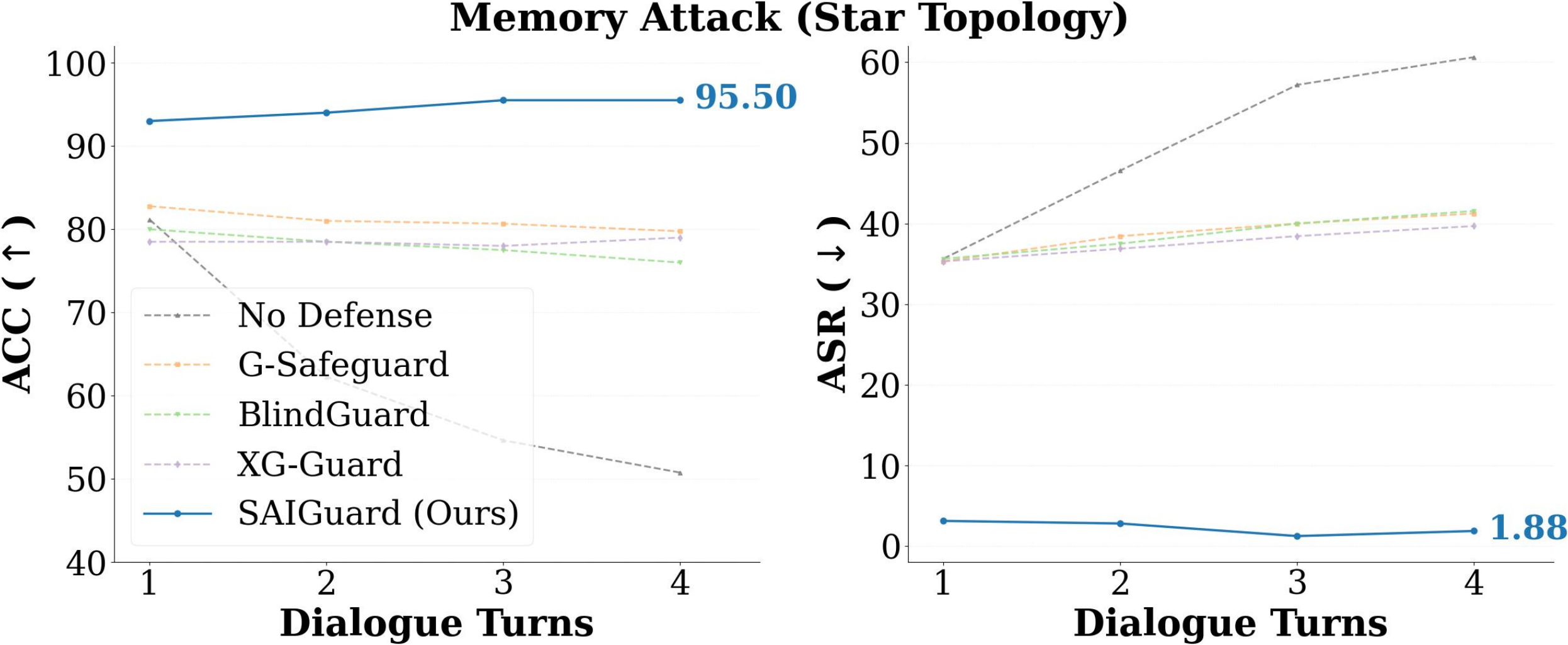}
 \vspace{-15pt}
    \caption{Impact of dialogue turns on defense efficacy. Evaluation of ACC ($\uparrow$) and ASR ($\downarrow$) under Memory Attack in Star Topology.}
    \label{fig:multi-turn2}
    \vspace{-15pt}
\end{figure}
To answer RQ1, we evaluate \ours\ from two perspectives: 
(1) overall defense effectiveness and task utility across different attack types, and 
(2) robustness and generalization across interaction topologies and backbone LLMs, and we have two observations: 
\textbf{\ding{182} \ours consistently defends against different attacks.}
Table~\ref{tab:main_results} shows that \ours\ achieves the best overall performance across four attack types, improving average ACC by $11.96\%$ and reducing average ASR by $67.47\%$ over the strongest baseline. Figure~\ref{fig:multi-turn2} further shows that under multi-turn memory attacks, baseline ASR increases as dialogue proceeds, while \ours\ maintains high ACC and low ASR in the Star topology. This confirms the effectiveness of proactive risk detection before real MAS execution, avoiding the utility degradation caused by post-hoc agent isolation.
\textbf{\ding{183} \ours\ is robust across topologies and generalizes across backbone models.}
Figure~\ref{fig:topology_generalization} evaluates the robustness and generalization of \ours.  As shown in Figure~\ref{fig:topology_generalization}(a), \ours\ achieves the lowest macro-average ASR across four topologies, indicating robustness to different interaction structures. Figure~\ref{fig:topology_generalization}(b) further shows that \ours\ consistently outperforms baseline defenses across Qwen3-30B-A3B and DeepSeek-V3.2, demonstrating its generalization across backbone LLMs. Appendix~\ref{app:additional experiments} provides detailed experiments.

\subsection{Scalability of \ours}
\begin{figure}[t]
    \centering
\includegraphics[width=1\linewidth]{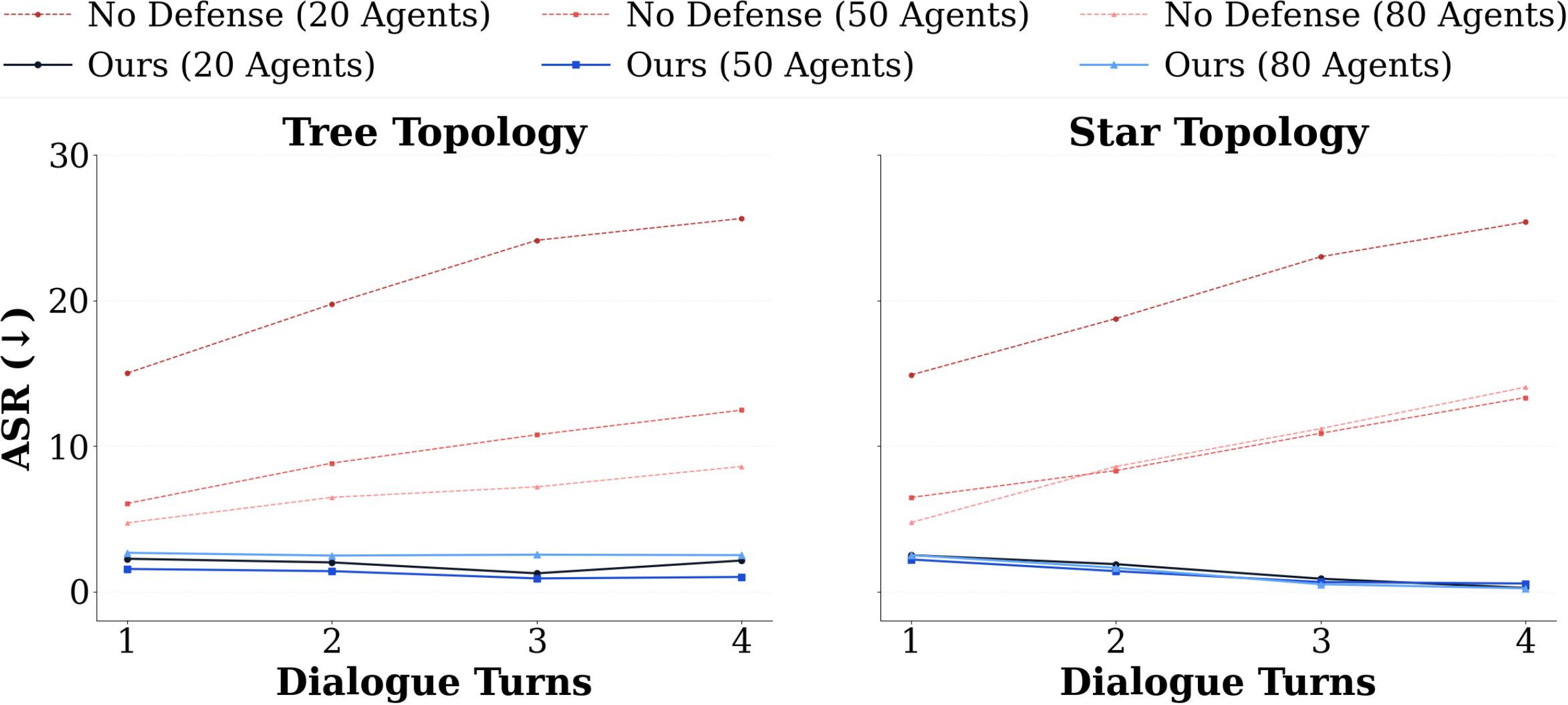}
\vspace{-10pt}
    \caption{Scalability analysis of \ours\ under varying agent numbers.}
    \label{fig:scalability}

\vspace{-15pt}

    % Compared to reactive baselines, \ours\ maintains minimal system risk regardless of topological complexity.
\end{figure}
To answer RQ2, we evaluate the scalability of \ours\ by varying MAS size ($20$, $50$, and $80$ agents) and topology (Chain, Tree, and Star). We observed \ding{184} \textbf{\ours\ generalizes well to larger-scale MAS with more agents.} As shown in Figure~\ref{fig:scalability}, ASR under no defense increases with dialogue turns, suggesting risk amplification in multi-round interactions. XG-Guard partially mitigates attacks but remains unstable across settings. In contrast, \ours\ consistently achieves the lowest ASR across all scales and topologies, even with $80$ agents. This demonstrates its strong scalability and topology-level generalization, as it suppresses attack propagation into MAS by simulating risks over the interaction graph before execution.
\subsection{Ablation Study}

To answer RQ3, we conduct an ablation study on the  PoisonRAG dataset, we observed \ding{185} \textbf{All components of \ours\ contribute to the final performance, especially the Communication-State Simulation component.} As reported in Table~\ref{tab:ablation_msmarco}, the full model achieves the best results across all metrics, reaching an F1-score of 98.68\%. The removal of the Communication-State Simulation(Com. Simulation) module causes the largest degradation, with F1 dropping by 2.73\%, highlighting its key role in modeling pre-execution communication and exposing latent adversarial cascades. Moreover, removing either the system-level or agent-level loss consistently hurts performance, validating the necessity of dual-level optimization. Finally, replacing MAD with Gaussian thresholding leads to inferior results, demonstrating the robustness of our median-based decision strategy under noisy multi-agent interactions.

\begin{table}[t]
\centering
\small
\begin{tabular}{lccc}
\toprule
\textbf{Architecture Variants} & \textbf{P (\%)} & \textbf{R (\%)} & \textbf{F1 (\%)} \\
\midrule
\ours (Ours) & \textbf{98.36} & \textbf{99.01} & \textbf{98.68} \\
\midrule
\quad w/o $\mathcal{L}_{\text{sys}}$ & 97.28 & 98.25 & 97.76 \\
\quad w/o $\mathcal{L}_{\text{agent}}$   & 97.14 & 98.55 & 97.84 \\
\quad w/o MAD                    & 97.39 & 98.35 & 97.87 \\
\quad w/o Com. Simulation                  & 95.24 & 96.68 & 95.95 \\
\bottomrule
\end{tabular}
 \vspace{-5pt}
\caption{Ablation study of \ours.}
\label{tab:ablation_msmarco}
   \vspace{-15pt}
\end{table}

\subsection{Threshold Sensitivity Coefficient Analysis}

\begin{wrapfigure}{r}{0.5\linewidth}
    \centering
    \vspace{-10pt}
    \includegraphics[width=1\linewidth]{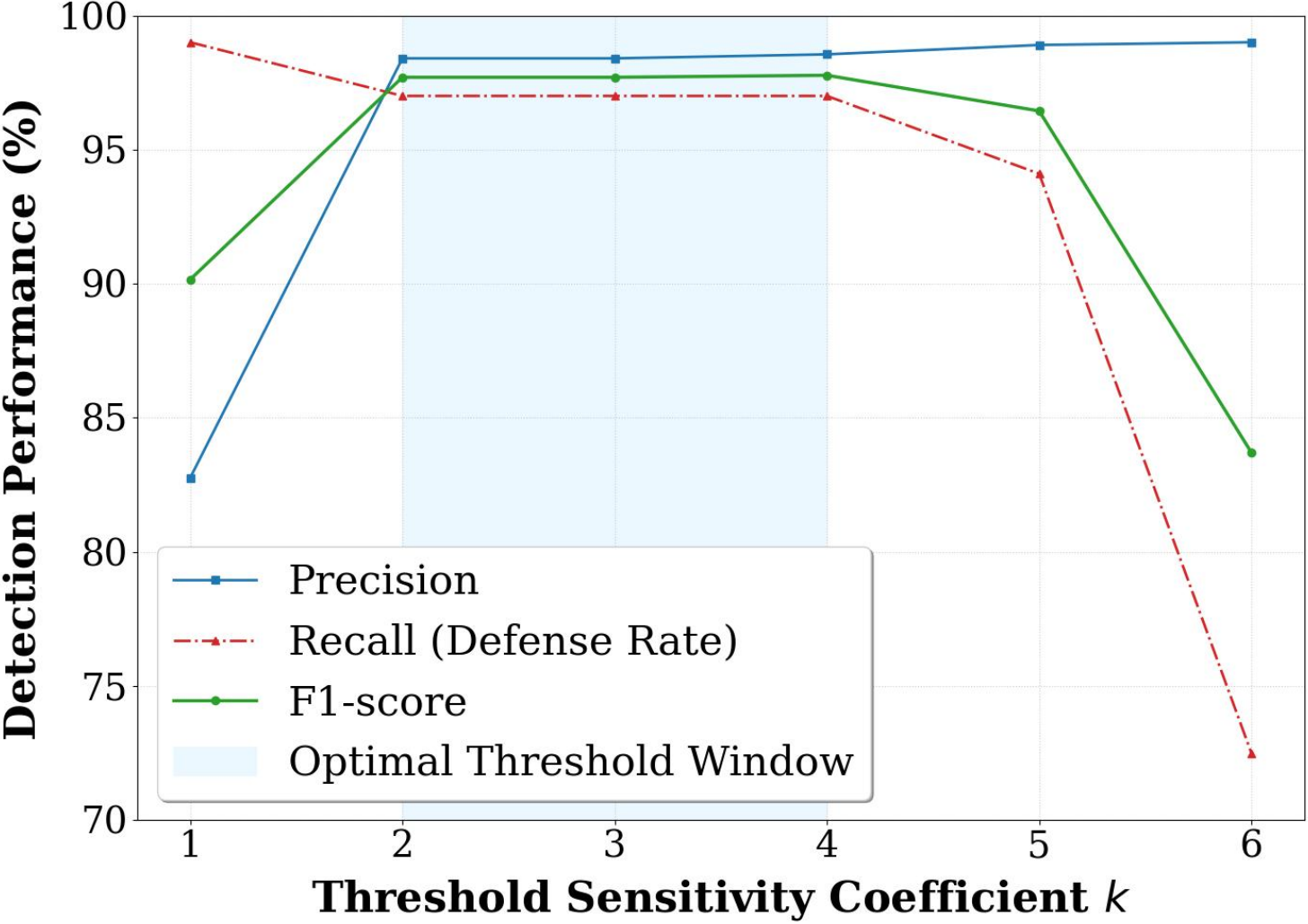}
     \vspace{-15pt}
    \caption{Sensitivity analysis of threshold $k$.}
    \label{fig:threshold}
    \vspace{-15pt}
\end{wrapfigure}

To answer RQ4, we evaluate the sensitivity of \ours\ to the threshold coefficient $k$ by varying $k$ from $1$ to $6$. 
We observed \ding{186} \textbf{\ours\ is robust to threshold variations and achieves stable detection performance within a practical threshold range.}
As shown in Figure~\ref{fig:threshold}, \ours\ achieves the most stable detection performance when $k$ falls within the optimal threshold window from $2$ to $4$. 
% In this range, precision, recall, and F1-score remain consistently high, indicating that \ours\ can maintain a good balance between detecting malicious inputs and avoiding false alarms. 
When $k$ is too small, the threshold becomes overly strict, leading to reduced precision due to excessive false positives. 
When $k$ is too large, the threshold becomes overly relaxed, causing recall and F1-score to drop as more malicious inputs are missed. 
This demonstrates that threshold calibration is important for balancing security and usability, and the default setting $k=3$ provides a robust performance.

\section{Conclusion}
\vspace{-5pt}
\label{sec:conclusion}
In this paper, we proposed \ours, a proactive defense framework for LLM multi-agent systems. 
\ours\ assesses incoming messages before execution through communication-state simulation, and detects risky messages by measuring local-global reconstruction deviations from benign communication patterns. 
Instead of isolating harmful agents after execution, \ours\ sanitizes or regenerates suspicious messages to mitigate risks while preserving collaboration. 
Experiments across diverse attacks, topologies, and backbone LLMs demonstrate that \ours\ effectively reduces attack success rates and maintains MAS utility.

\FloatBarrier 
\section*{Limitations}
Although \ours demonstrates effective proactive defense for LLM-based MAS, it currently focuses on textual communication among agents. However, real-world MAS may also involve multimodal inputs, such as images, audio, or videos. As a result, \ours may not fully capture risks hidden in non-textual modalities, including visual prompt injection and misleading visual content.
Future work can extend \ours to multimodal MAS by incorporating multimodal encoders into the communication-state simulation module. These encoders can map visual, audio, and structured inputs into a unified interaction graph together with textual messages. The intervention module can also be extended to support modality-aware remediation, such as visual-content verification, and multimodal response regeneration.

% \section*{Acknowledgments}

\bibliography{custom}

\newpage

\appendix
\section{Related Work}
\paragraph{LLM-based Multi-agent Systems.} Recent advances in LLM-based MAS have demonstrated strong capabilities in general task-solving~\cite{li2024survey,leong2025amas,chen2025optima,cemri2026multi}. Compared with single-agent systems~\cite{zhang2024agent,luo2026agentauditor}, MAS improve performance on complex tasks by enabling agents with diverse roles and expertise to collaborate through structured communication~\cite{wang2025anymac}. Prior work has explored various coordination mechanisms, including sequential reasoning~\cite{zhou2025reso}, and debate-based collaboration~\cite{choi2026debate}. However, the communication mechanisms that enhance collaboration also expand the attack surface~\cite{yu2025netsafe}, allowing malicious information to propagate across agents through multi-round interactions~\cite{zhou2025guardian}.

\paragraph{Security Risks and Defenses in LLM-based MAS.} LLM-based MAS face various security threats, including prompt injection~\cite{wang2025webinject}, tool attacks~\cite{zhan2024injecagent}, memory poisoning~\cite{nazary2025poison,chen2024agentpoison}, communication hijacking~\cite{he2025red}. Moreover, this attacks  can propagate through inter-agent communication~\cite{zhou2025guardian}, causing cascading failures and amplifying systemic security risks. Existing defenses mainly detect abnormal agents or unsafe behaviors after execution. Such as, G-Safeguard~\cite{wang2025g}, BlindGuard~\cite{miao2025blindguard} and XG-Guard~\cite{pan2025explainable}, further use GNNs, contrastive, or trust modeling learning to identify harmful agents. Despite their effectiveness, most existing methods are reactive and rely on post-execution observations, which may lead to irreversible damage and disrupt collaboration through agent isolation. In contrast, \ours\ proactively simulates message propagation over the MAS interaction graph and detects harmful inputs before execution.

\section{Parallelized Algorithm}
\label{app:algorithm}
\begin{algorithm*}[t]
\small
\caption{Parallelized \ours Procedure}
\label{alg:ours}
\begin{algorithmic}[1]
\Require MAS graph $\mathcal{G}=(\mathcal{V},\mathcal{E},\mathbf{A})$, benign traces $\mathcal{D}_{b}$, incoming message $M_i \in \mathcal{M}$, propagation depth $L$
\Ensure Anomaly decision and intervention action

\Statex \textbf{Stage 1: Learning Normal MAS Patterns}
\For{each benign trace in $\mathcal{D}_{b}$}
    \State Encode each agent $v_i$ into $h_i$ \Comment{Agent-level MAS embedding}
    \State Initialize $\mathbf{x}_i^{(0)}$ with $h_i$ and $M_i$ \Comment{Message-aware state initialization}
    \For{$l=1$ to $L$}
        \State Update agent states by GNN message passing \Comment{Simulate multi-hop propagation}
    \EndFor
    \State Aggregate agent states into $\mathbf{g}$ \Comment{Obtain global MAS state}
    \State Reconstruct agent and system states with $D_\theta$ \Comment{Benign-pattern reconstruction}
    \State Optimize $D_\theta$ by minimizing $\mathcal{L}_{rec}$ \Comment{Learn normal MAS patterns}
\EndFor

\Statex \textbf{Stage 2: Reconstruction-based Threshold Estimation}
\State Collect agent-level reconstruction errors into $S_{agent}$ \Comment{Agent-level benign errors}
\State Collect system-level reconstruction errors into $S_{sys}$ \Comment{System-level benign errors}
\State Compute $\tau_{agent}=\tau(S_{agent})$ and $\tau_{sys}=\tau(S_{sys})$ \Comment{Calibrate anomaly thresholds}

\Statex \textbf{Stage 3: Local-Global Deviation Intervention}
\State Inject incoming message $M_i$ into its recipient agent \Comment{Pre-execution risk assessment}
\State Simulate propagation and obtain $\{\tilde{\mathbf{x}}_i^{(L)}\}_{i=1}^{N}$ and $\tilde{\mathbf{g}}$ \Comment{Estimate potential impact}
\State Reconstruct simulated states using $D_\theta$ \Comment{Compare with benign patterns}
\State Compute $\{e_i^n\}_{i=1}^{N}$ and $e^{\mathbf{g}}$ \Comment{Local-global anomaly scoring}
\State Determine $\mathcal{A}(\mathcal{M})$ using $\tau_{agent}$ and $\tau_{sys}$ \Comment{Deviation detection}
\If{$\mathcal{A}(\mathcal{M})=1$}
    \State $\mathcal{B}(M_i)
    =
    \mathbb{I}
    \left[
        e_j^{n} > \tau_{agent}
    \right],
    \quad v_j \in \mathcal{V},\quad M_i \in \mathcal{M}.$
    \If{$\mathcal{B}(M_i)=1$} 
    \State Apply targeted intervention before execution \Comment{Prevent risky propagation}
    \State block, or regenerate risky communication \Comment{Mitigate detected risk}
    \Else
    \State Allow $M_i$ to enter the MAS \Comment{Benign message execution}
    \EndIf

\Else
    \State Allow $\mathcal{M}$ to enter the MAS \Comment{Benign message execution}
\EndIf
\end{algorithmic}
\end{algorithm*}

% \begin{figure}[t]
%     \centering
% \includegraphics[width=1\linewidth]{latex/figure/Intervention.pdf}
%     \caption{Intervention strategy of \ours.}
%     \label{fig:intervention}

%     % Compared to reactive baselines, \ours\ maintains minimal system risk regardless of topological complexity.
% \end{figure}
Algorithm~\ref{alg:ours} presents the complete parallelized workflow of \ours, which includes training, threshold estimation, and inference-time intervention. 
In the training stage, \ours first represents the MAS as an interaction graph, where agents are treated as nodes and their communication links define the graph structure. 
For each benign execution trace, the agent states are initialized by combining agent representations with message embeddings, and an $L$-layer GNN is used to simulate how messages propagate across the MAS. 
After multi-round propagation, the agent representations are aggregated into a global MAS representation, which captures the overall communication state of the system. 
A decoder is then trained to reconstruct both agent-level states and the global system state. 
By minimizing the joint reconstruction loss on benign traces, \ours learns the normal communication and state-transition patterns of the MAS.

After training, \ours estimates anomaly thresholds from reconstruction errors observed under benign executions. 
Specifically, agent-level errors are collected to characterize normal local deviations, while system-level errors are collected to characterize normal global deviations. 
These benign error distributions are then used to compute robust thresholds for later anomaly detection:
\begin{equation}
\tau_{agent} = T(S_{agent}), 
\quad 
\tau_{sys} = T(S_{sys}),
\end{equation}
where \(S_{agent}\), and \(S_{sys}\) denote the benign agent-level and system-level reconstruction error distributions, respectively.

During inference, before incoming message is executed, \ours injects it into the simulated recipient agent and estimates its potential propagation effect over the MAS graph. 
After simulating the propagation of the message \(\mathcal{M}\) over the MAS graph, \ours reconstructs the resulting agent and system states with the benign-trained decoder and computes their reconstruction errors. 
It first makes a system-level decision by comparing the local and global deviations against calibrated thresholds:
\begin{equation}
\mathcal{A}(\mathcal{M})
=
\mathbb{I}
\left[
    e_\mathbf{g} > \tau_{sys}
    \;\vee\;
    \max_{v_j \in \mathcal{V}} e_j > \tau_{agent}
\right].
\end{equation}
If \(\mathcal{A}(\mathcal{M})=0\), all messages are allowed to enter the MAS. 
Otherwise, \ours performs localization by evaluating each \(M_i \in \mathcal{M}\) with agent-level deviations.

Messages satisfying $B(M_i)=1$ are identified as risky and subjected to targeted intervention before execution:
\begin{equation}
\mathcal{B}(M_i)
=
\mathbb{I}
\left[
    e_i^{n} > \tau_n
\right].
\end{equation}
Specifically, \ours adopts a source-aware intervention strategy for different attack scenarios. 
In agent-target attacks, the abnormal message originates from an external malicious source and attempts to enter the multi-agent system through a target agent. 
\ours therefore performs boundary-level blocking before the message is propagated into the MAS. 
In communication-target attacks, the sender agent remains benign while the transmitted message is hijacked or tampered with. 
In this case, \ours discards the compromised message but does not isolate the benign sender. 
Instead, it requests the benign sender agent to regenerate the communication content. 
This design enables \ours to prevent external injection while preserving benign collaboration.

% , including sanitization, blocking, or regeneration, while benign messages are allowed to proceed.

% The incoming message first triggers a local-global deviation alarm if either the system-level error or the maximum agent-level error exceeds the calibrated threshold:
% \begin{equation}
% \mathcal{G}(M_i)
% =
% \mathbb{I}
% \left[
%     e^{\mathbf{g}} > \tau_{\mathbf{g}}
%     \;\vee\;
%     \max_{v_j \in \mathcal{V}} e_j^{n} > \tau_n
% \right].
% \end{equation}
% Once the alarm is triggered, \ours further localizes affected agents using agent-level deviations:
% \begin{equation}
% \mathcal{B}_j(M_i)
% =
% \mathcal{G}(M_i)
% \cdot
% \mathbb{I}
% \left[
%     e_j^{n} > \tau_n
% \right],
% \quad v_j \in \mathcal{V}.
% \end{equation}
% Here, $\mathcal{B}_j(M_i)=1$ indicates that agent $v_j$ is selected for targeted intervention. 
% The system-level deviation score provides a global alarm, while the agent-level deviation scores determine which agents should be sanitized, blocked, or regenerated.

\section{Detailed Experiment Setups}
\subsection{Scientific Artifacts}
All external artifacts are used solely for research and evaluation purposes, consistent with their licenses and terms of use. The generated multi-agent interaction traces and detection datasets are intended for research on safety, robustness, and defense in multi-agent LLM systems, and should not be used to deploy or facilitate harmful behavior. We removed or replaced identifiable placeholders such as names and email addresses with anonymized examples (e.g., anonymized User and anonymized@example.com). Our generated multi-agent traces are synthetic and do not contain real user identities. Potentially harmful or adversarial prompts are used only for research on safety and defense, and are released with restrictions for research purposes only.
\subsection{Baselines Details}
\label{suba:baseline}
For fair comparison, all MAS defense baselines are evaluated under the same MAS setting, with identical agent numbers, roles, communication topology. We compare \ours\ with six representative MAS anomaly detection baselines, including MAS-specific safeguarding methods and general graph anomaly detection methods adapted to MAS communication graphs.

\textbf{G-Safeguard}~\cite{wang2025g} is a supervised safeguarding framework designed for LLM-based multi-agent systems. 
It represents MAS interactions as a multi-agent utterance graph and applies graph neural networks to detect harmful agents from the communication topology. 
After detecting harmful agents, G-Safeguard isolate harmful agents to mitigate the propagation of malicious information. It is a strong MAS-specific baseline for evaluating graph-based security defenses.

\textbf{DOMINANT}~\cite{ding2019deep} is a classical deep graph anomaly detection method for attributed networks. 
It uses a graph convolutional encoder to learn node embeddings from both graph structure and node attributes, and then reconstructs the graph through decoder modules. 
Nodes with large reconstruction errors are treated as anomalies. 
In our MAS setting, we adapt DOMINANT by treating agent outputs as nodes and inter-agent communication links as edges. 
This baseline evaluates whether reconstruction-based graph anomaly detection can identify abnormal agents in MAS communication graphs.

\textbf{PREM}~\cite{pan2023prem} is an efficient node-level graph anomaly detection method based on preprocessing and ego-neighbor matching. Unlike message-passing-heavy GNN methods, PREM reduces training complexity by using a lightweight preprocessing module and a matching objective between each node and its local neighborhood. 
Anomalies are detected based on inconsistencies between a node and its surrounding context. 
We adapt PREM to MAS anomaly detection by applying its ego-neighbor matching mechanism to agent communication graphs, where abnormal agents are expected to deviate from their neighboring agents in representation space.

\textbf{TAM}~\cite{qiao2023truncated} is an unsupervised graph anomaly detection method based on one-class homophily modeling. 
It assumes that normal nodes tend to have stronger affinity with their neighbors, while anomalous nodes show weaker neighborhood consistency. 
TAM maximizes local node affinity and iteratively removes potentially non-homophilous edges to reduce the influence of abnormal connections. 
In our experiments, we use TAM to detect agents whose communication representations are weakly aligned with their neighboring agents, making it a suitable baseline for testing local consistency-based anomaly detection in MAS.

\textbf{BlindGuard}~\cite{miao2025blindguard} is an unsupervised defense framework for LLM-based MAS under unknown attacks. 
Instead of relying on labeled harmful agents, BlindGuard learns from normal agent behaviors and uses a hierarchical agent encoder to capture individual, neighborhood, and global interaction patterns. 
It further introduces a corruption-guided detector with contrastive learning to distinguish abnormal agents from normal ones. 

\textbf{XG-Guard}~\cite{pan2025explainable} is an explainable and fine-grained MAS safeguarding method based on bi-level graph anomaly detection. 
It models both sentence-level and token-level information of agent responses, allowing the detector to capture coarse semantic anomalies as well as fine-grained lexical cues. 
XG-Guard further provides token-level explanations for detected malicious behaviors, improving the interpretability of MAS defense. 
\subsection{Metrics Details}
\label{app:metrics}

We evaluate MAS defense methods from two complementary perspectives: 
system-level MAS behavior and hazardous-message detection performance. 
Task Accuracy (ACC) and Attack Success Rate (ASR) are used to measure system-level MAS behavior, while Precision, Recall, and F1-score are used to evaluate whether \ours can correctly identify hazardous input messages.

\paragraph{System-level Task Accuracy (ACC).}
ACC measures whether the MAS can complete the user task. 
Unlike single-agent evaluation, MAS outputs may contain responses from multiple agents in each communication round. 
Therefore, we evaluate task correctness at the system level. 
For each test instance, we extract the answer from each agent response using the required answer format, e.g., \texttt{<ANSWER>: A} for multiple-choice tasks or \texttt{<ANSWER>: 42} for numerical reasoning tasks. 
Given the ground-truth answer, an agent response is considered correct if the extracted answer matches the ground truth.

For a MAS with $N$ agents, we regard the system as correct in a given communication round if at least half of the agents provide the correct answer. 
Formally, for instance $q$ at round $t$, let $a_{q,t,i}$ be the answer extracted from agent $i$, and let $y_q$ be the ground-truth answer. 
The system-level correctness is defined as:
\begin{equation}
C_{q,t}
=
\mathbb{I}
\left[
\sum_{i=1}^{N}
\mathbb{I}(a_{q,t,i}=y_q)
\geq
\frac{N}{2}
\right].
\end{equation}
Then, ACC at round $t$ is computed as:
\begin{equation}
\mathrm{ACC}_{t}
=
\frac{1}{|\mathcal{D}|}
\sum_{q\in\mathcal{D}}
C_{q,t},
\end{equation}
where $\mathcal{D}$ denotes the evaluation set. 
This metric reflects whether the MAS, as a whole, can preserve task utility under attack. 
In our implementation, ACC is computed for each communication round, which allows us to analyze how task performance changes as agents exchange more messages. 
% For MMLU and GSM8K, the extracted answers are evaluated as choice-based and number-based answers, respectively. 
% The evaluation code computes MAS-level correctness by checking whether the number of correct agent responses reaches at least half of all agents. 
For MMLU~\cite{hendrycks2020measuring} and GSM8K~\cite{cobbe2021training}, we extract answers from agent responses and evaluate them as choice-based and number-based answers, respectively. 
For PoisonRAG~\cite{nazary2025poison}, where answers are open-ended, we use an LLM-based evaluator to judge whether each agent response is semantically consistent with the reference answer. 
For InjecAgent~\cite{zhan2024injecagent}, we use an LLM-based evaluator to determine whether each agent output successfully answers the original user instruction using the trusted tool observation. 
MAS-level ACC is then computed by aggregating agent-level correctness within each communication round: for MMLU~\cite{hendrycks2020measuring} and GSM8K~\cite{cobbe2021training}, the MAS is considered correct if at least half of the agents produce the correct answer, while for InjecAgent, the MAS is considered successful if the system-level success criterion is satisfied under the LLM-based evaluator.

\paragraph{Attack Success Rate (ASR).}
ASR measures the extent to which an attack successfully compromises the MAS behavior. 
A lower ASR indicates better defense effectiveness. 
Since different attacks in different forms, we compute ASR according to the attack objective of each benchmark.

For answer-based attacks, such as GSM8K~\cite{cobbe2021training}, and MMLU~\cite{hendrycks2020measuring}, ASR measures whether the MAS output is shifted away from the correct answer or toward the attacker-specified wrong answer. 
For each communication round, we extract answers from agent responses and compare them with the ground-truth answer or attacker target. 
If the attack causes the MAS to produce an incorrect or attacker-desired answer, it is counted as a successful attack. 
This evaluates whether the defense can prevent malicious messages from corrupting the system-level task result.

For tool-use attacks, such as InjecAgent-style attacks~\cite{zhan2024injecagent}, ASR measures whether non-attacker agents are induced to execute the attacker-specified tool or follow the attacker instruction. 
The evaluator parses the agent output in ReAct format, checks whether the target malicious tool is invoked, and marks the attack as successful when a non-attacker agent performs the attacker-desired action. 
The ASR is then computed over MAS communication rounds:
\begin{equation}
\mathrm{ASR}_{t}
=
\frac{1}{|\mathcal{D}|}
\sum_{q\in\mathcal{D}}
\mathbb{I}
\left[
\text{AttackSucceeds}(q,t)
\right].
\end{equation}
In our implementation, ASR is also computed for each communication round, enabling us to observe whether attacks are amplified or suppressed during multi-round MAS communication. 

For PoisonRAG-style tasks~\cite{nazary2025poison}, attack success is determined by whether the extracted answer contains the attacker-specified incorrect answer. 

\paragraph{Precision, Recall, and F1-score.}
While ACC and ASR evaluate the final MAS behavior, Precision, Recall, and F1-score evaluate the detection ability of a defense method at the hazardous-message level. Specifically, Precision measures how many detected hazardous messages are truly hazardous. Recall measures how many hazardous messages are successfully detected. F1-score is the harmonic mean of Precision and Recall. 

These detection metrics complement ACC and ASR. 
High ACC indicates that the defense preserves benign task utility, while low ASR indicates that it effectively suppresses malicious influence at the MAS level. 
Precision, Recall, and F1-score further show whether the defense accurately identifies hazardous messages without over-blocking benign communication.

\subsection{Implementation Details}
\label{app:implementation}

We provide more implementation details of \ours in this section, including MAS construction, feature extraction, model training, threshold calibration, and inference-time intervention.

\paragraph{MAS Construction.}
For all experiments, we construct a multi-agent system with $8$ agents and $3$ attack instances unless otherwise specified. 
Each MAS is represented as a directed communication graph, where nodes denote agents and edges denote directed communication links. 
To evaluate topological robustness, we instantiate four types of MAS topologies: Chain, Tree, Star, and Random. 
For the random topology, we generate directed adjacency matrices with edge sparsity $0.5$ and remove self-loops. 
For Chain, Tree, and Star topologies, we use fixed topology templates with the same number of agents. 
Across all baselines and \ours, we keep the MAS configuration identical, including the number of agents, attacker ratio, task inputs, communication topology, and dialogue turns.

\paragraph{Datasets and Attack Settings.}
We evaluate \ours on four datasets: MMLU~\cite{hendrycks2020measuring}, PoisonRAG~\cite{nazary2025poison}, InjecAgent~\cite{zhan2024injecagent}, and GSM8K~\cite{cobbe2021training}. 
For MMLU~\cite{hendrycks2020measuring}, agents collaboratively solve multiple-choice questions, while attacker agents attempt to steer the discussion toward an incorrect answer. 
For PoisonRAG~\cite{nazary2025poison}, agents answer open-ended retrieval-based questions, and attacks are introduced through misleading or poisoned contexts. 
For InjecAgent~\cite{zhan2024injecagent}, agents operate in a tool-integrated environment, where attacks attempt to induce non-attacker agents to follow hidden malicious instructions or invoke attacker-specified tools. 
For GSM8K~\cite{cobbe2021training}, agents solve mathematical reasoning problems, while attackers attempt to disrupt the reasoning process and mislead other agents. 
Each sample is executed for one initial response round followed by $3$ additional communication rounds, producing multi-turn MAS communication traces for both defended and undefended settings.

\paragraph{Backbone LLMs.}
Unless otherwise specified, we use GPT-4o-mini as the default backbone LLM. 
To evaluate generalization across backbone models, we additionally conduct experiments with DeepSeek-V3~\cite{liu2024deepseek}, DeepSeek-V3.2~\cite{liu2025deepseek}, and Qwen-30B-A3B~\cite{yang2025qwen3}. 
All compared methods are evaluated under the same backbone LLM setting for fair comparison.

\paragraph{Model Training.}
All experiments were conducted on a single NVIDIA L40. We use Qwen3-Embedding-0.6B to encode agent states and communication contents into dense representations. Then, we train the system reconstruction model only on benign MAS execution traces. 
The model takes agent features and the MAS adjacency matrix as input, and reconstructs both agent-level states and the global MAS state. 
We use a hidden dimension of $1024$, $2$ graph layers, and dropout rate $0.2$. 
The model is trained for $20$ epochs with Adam optimizer, learning rate $1\times10^{-3}$, weight decay $2\times10^{-4}$, and batch size $32$. 
We apply a cosine annealing learning-rate scheduler with $T_{\max}=10$ and minimum learning rate $1\times10^{-5}$. 

\paragraph{Threshold Calibration.}
After training, we collect reconstruction errors on benign training traces to calibrate anomaly thresholds. 
Specifically, we compute node-level reconstruction scores for agent-level deviations and graph-level reconstruction scores for system-level deviations. 
Then, we compute thresholds from benign reconstruction error distributions:
\begin{equation}
\tau(S)=\operatorname{median}(S)+k\cdot 1.4826\cdot \operatorname{MAD}(S),
\end{equation}
where $S$ denotes a benign error set and $k$ controls detection sensitivity. 
In our experiments, we set $k=3$ following the common three-sigma rule. 
This produces an agent-level threshold $\tau_n$ and a system-level threshold $\tau_{sys}$, which are used during inference to detect abnormal deviations.

\begin{figure*}[t]
    \centering
\includegraphics[width=1\linewidth]{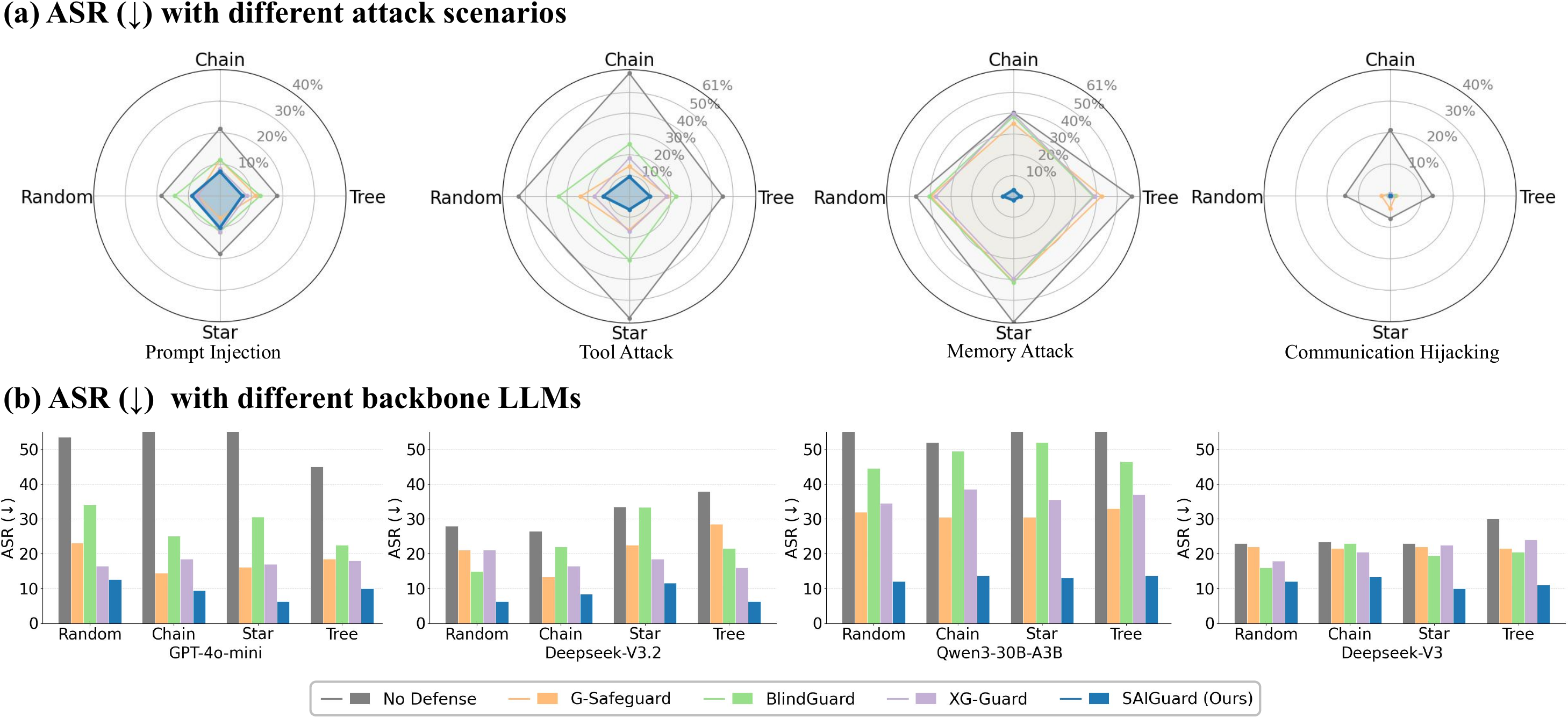}
    \caption{ASR comparison across communication topologies and backbone LLMs.}
    \label{fig:asr_app}

    % Compared to reactive baselines, \ours\ maintains minimal system risk regardless of topological complexity.
\end{figure*}

\subsection{Detailed analysis under multi-turn dialogues.}
\begin{figure*}[t]
    \centering
\includegraphics[width=1\linewidth]{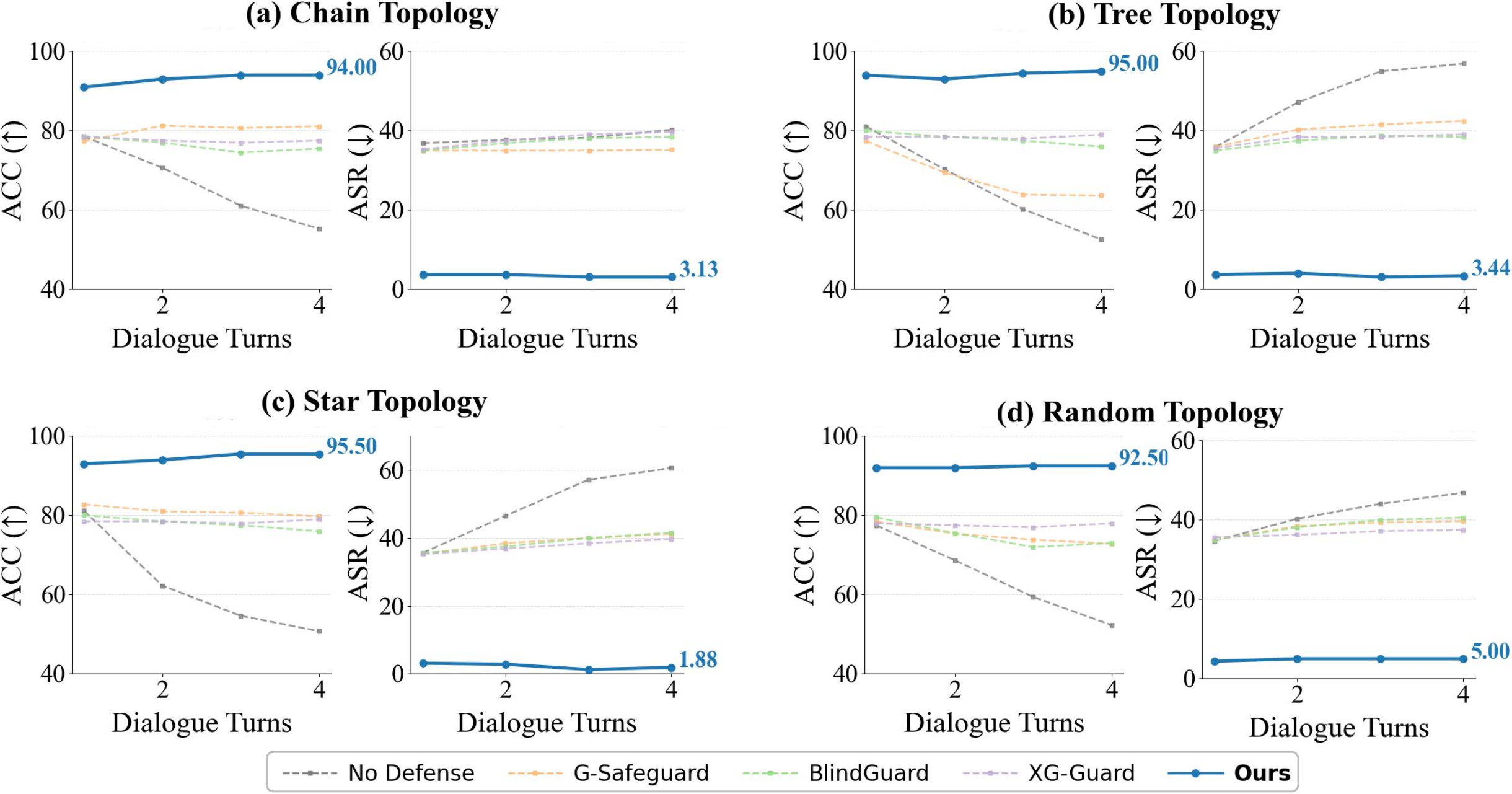}
    \caption{Impact of dialogue turns on defense efficacy.}
    \label{fig:multi-turn_app}

    % Compared to reactive baselines, \ours\ maintains minimal system risk regardless of topological complexity.
\end{figure*}

\paragraph{Inference-time Defense.}
During inference, \ours first simulates the potential impact of incoming messages or agent states under the current MAS topology. 
The trained reconstruction model then computes both agent-level anomaly scores and a system-level anomaly score. 
A message or MAS state is flagged as anomalous if either the system-level score exceeds $\tau_{sys}$ or any agent-level score exceeds $\tau_{agent}$. 
For initial-system or agent-level attacks, \ours replaces detected malicious prompts with benign prompts. 
For communication-level attacks, \ours removes corrupted communication message, and regenerates clean messages from the agents. 
The final defended MAS trace is then saved and evaluated using the same task-level and security-level metrics as all baselines.

\section{Additional Experiments}
\label{app:additional experiments}
\subsection{Comprehensive Evaluation}
\label{app:e1}
\begin{table*}[t]
\centering

\small
\resizebox{\textwidth}{!}{
\begin{tabular}{llcccccccc|cc}
\toprule
\multirow{2}{*}{\textbf{Topology}} & \multirow{2}{*}{\textbf{Defense}} 
& \multicolumn{2}{c}{\textbf{Prompt Injection}} 
& \multicolumn{2}{c}{\textbf{Tool Attack}} 
& \multicolumn{2}{c}{\textbf{Memory Poisoning}} 
& \multicolumn{2}{c}{\textbf{Com. Hijacking}} 
& \multicolumn{2}{|c}{\textbf{Avg.}} \\ 
\cmidrule(lr){3-4} \cmidrule(lr){5-6} \cmidrule(lr){7-8} \cmidrule(lr){9-10} \cmidrule(lr){11-12}
& & ACC $\uparrow$ & ASR $\downarrow$ 
& ACC $\uparrow$ & ASR $\downarrow$ 
& ACC $\uparrow$ & ASR $\downarrow$ 
& ACC $\uparrow$ & ASR $\downarrow$ 
& ACC $\uparrow$ & ASR $\downarrow$ \\ 
\midrule

\multirow{8}{*}{Chain} 
& No Defense  & 85.02 & 21.82 & 30.45 & 59.27 & 55.28 & 40.25 & 70.28 & 20.94 & 60.26 & 35.57 \\
& Dominant    & 85.21 & 18.57 & 40.00 & 58.00 & 61.25 & 39.44 & 83.21 & 3.75  & 67.42 & 29.94 \\
& PREM        & 85.32 & 17.43 & 47.59 & 36.00 & 61.46 & 39.19 & 83.54 & 3.13  & 69.48 & 23.94 \\
& TAM         & 85.35 & 16.57 & 46.67 & 47.00 & 70.64 & 35.31 & 82.58 & 4.38  & 71.31 & 25.82 \\
& G-Safeguard & 85.78 & 11.39 & 67.52 & 14.55 & 81.09 & 35.21 & 85.64 & 0.62  & 80.01 & 15.44 \\
& BlindGuard  & 85.56 & 11.42 & 57.54 & 25.13 & 75.50 & 38.44 & \textbf{87.52} & 0.64  & 76.53 & 18.91 \\
& XG-Guard    & 87.47 & 8.57  & 67.58 & 18.50 & 77.50 & 39.68 & 85.10 & 0.63  & 79.41 & 16.85 \\
& \textbf{\ours} & \textbf{90.79} & \textbf{7.69} & \textbf{86.78} & \textbf{9.47} & \textbf{94.00} & \textbf{3.13} & 87.51 & \textbf{0.00} & \textbf{89.79} & \textbf{5.07} \\

\midrule
\multirow{8}{*}{Tree} 
& No Defense  & 82.57 & 18.18 & 32.52 & 45.01 & 52.59 & 56.88 & 80.03 & 13.44 & 61.93 & 33.38 \\
& Dominant    & 82.55 & 18.05 & 40.00 & 44.50 & 58.89 & 50.63 & 83.19 & 3.94  & 66.16 & 29.28 \\
& PREM        & 83.04 & 17.15 & 41.38 & 42.05 & 61.18 & 49.69 & 83.52 & 3.44  & 67.28 & 28.08 \\
& TAM         & 84.11 & 16.57 & 28.57 & 41.50 & 62.42 & 44.06 & 82.47 & 5.94  & 64.39 & 27.02 \\
& G-Safeguard & 85.87 & 11.48 & 70.02 & 18.55 & 77.41 & 42.45 & 85.71 & 1.79  & 79.75 & 18.57 \\
& BlindGuard  & 85.64 & 12.86 & 55.09 & 22.58 & 76.11 & 38.44 & 87.52 & 1.56  & 76.09 & 18.86 \\
& XG-Guard    & 87.54 & 8.57  & 67.51 & 18.06 & 79.27 & 39.09 & \textbf{90.00} & 0.60  & 81.08 & 16.58 \\
& \textbf{\ours} & \textbf{87.71} & \textbf{7.14} & \textbf{89.47} & \textbf{10.04} & \textbf{95.02} & \textbf{3.44} & 87.59 & \textbf{0.00} & \textbf{89.95} & \textbf{5.16} \\

\midrule
\multirow{8}{*}{Star} 
& No Defense  & 85.21 & 18.51 & 45.12 & 58.49 & 50.77 & 60.63 & 85.11 & 7.19  & 66.55 & 36.21 \\
& Dominant    & 85.22 & 17.41 & 50.77 & 57.18 & 52.46 & 59.68 & 83.17 & 3.98  & 67.91 & 34.56 \\
& PREM        & 85.17 & 16.55 & 54.44 & 31.50 & 60.85 & 54.62 & 83.28 & 3.31  & 70.94 & 26.50 \\
& TAM         & 85.24 & 15.48 & 52.48 & 46.76 & 68.47 & 49.02 & 84.74 & 2.50  & 72.73 & 28.44 \\
& G-Safeguard & \textbf{87.53} & \textbf{7.09} & 70.03 & 16.16 & 79.76 & 41.25 & 85.57 & 4.06  & 80.72 & 17.14 \\
& BlindGuard  & 85.24 & 11.43 & 55.10 & 30.59 & 76.00 & 41.56 & 87.51 & 0.31  & 75.96 & 20.97 \\
& XG-Guard    & 85.21 & 11.47 & 70.01 & 17.00 & 79.00 & 39.69 & 82.53 & 0.31  & 79.19 & 17.12 \\
& \textbf{\ours} & 85.78 & 10.19 & \textbf{89.47} & \textbf{6.32} & \textbf{95.50} & \textbf{1.88} & \textbf{90.21} & \textbf{0.00} & \textbf{90.24} & \textbf{4.60} \\

\midrule
\multirow{8}{*}{Random} 
& No Defense  & 85.53 & 18.54 & 52.51 & 53.47 & 52.29 & 46.88 & 75.18 & 14.38 & 66.38 & 33.32 \\
& Dominant    & 85.29 & 17.43 & 55.77 & 49.50 & 58.31 & 45.31 & 83.00 & 3.56  & 70.59 & 28.95 \\
& PREM        & 85.31 & 16.86 & 57.62 & 45.64 & 64.84 & 43.75 & 83.31 & 3.88  & 72.77 & 27.53 \\
& TAM         & 85.05 & 18.01 & 58.87 & 45.00 & 65.07 & 43.75 & 82.89 & 4.50  & 72.97 & 27.82 \\
& G-Safeguard & \textbf{90.00} & \textbf{7.54} & 60.07 & 23.10 & 72.82 & 39.69 & 87.53 & 2.81  & 77.61 & 18.29 \\
& BlindGuard  & 87.51 & 14.29 & 52.25 & 34.11 & 73.00 & 40.63 & 90.01 & 0.63  & 75.69 & 22.42 \\
& XG-Guard    & 87.59 & 7.61  & 67.50 & 16.49 & 78.00 & 37.50 & 90.31 & 0.31  & 80.85 & 15.48 \\
& \textbf{\ours} & 87.53 & 8.93 & \textbf{86.84} & \textbf{12.63} & \textbf{92.50} & \textbf{5.00} & \textbf{90.51} & \textbf{0.00} & \textbf{89.35} & \textbf{6.64} \\

\bottomrule
\end{tabular}
}
\caption{Comprehensive experimental results across all topologies and attack Types. We report Task Accuracy (ACC \%) and Attack Success Rate (ASR \%) for each configuration. The ``Avg'' column represents the mean performance across the four evaluated attack types within each topology. \textbf{Bold} indicates the best performance (highest ACC and lowest ASR) among all defense methods.}
\label{tab:full_results_appendix}
\end{table*}
Table~\ref{tab:full_results_appendix} provides the complete results across four MAS topologies, including Chain, Tree, Star, and Random. Overall, \ours achieves the best average performance under all topologies, obtaining 89.79 ACC / 5.07 ASR on Chain, 89.95 ACC / 5.16 ASR on Tree, 90.24 ACC / 4.60 ASR on Star, and 89.35 ACC / 6.64 ASR on Random. Compared with the strongest baseline in each topology, \ours improves average ACC by 9.78, 8.87, 9.52, and 8.50 points, respectively, while reducing average ASR by 10.37, 11.42, 12.52, and 8.84 points. These results show that the advantage of \ours is not limited to a specific interaction structure, but remains consistent across both regular and random communication patterns.

The detailed attack-wise results further show that \ours is especially effective against attacks that propagate through MAS communication. For communication hijacking, \ours reduces ASR to 0.00 under all four topologies, indicating that risky inter-agent messages can be intercepted before affecting downstream agents. For memory poisoning, \ours keeps ASR between 1.88 and 5.00 while maintaining ACC above 92.50 across all topologies, substantially outperforming existing defenses. For tool attacks, \ours also achieves consistently high ACC, with much lower ASR than baselines. These results suggest that communication-state simulation helps \ours capture the potential system-level impact of malicious messages before they spread.

We also observe that \ours does not always achieve the best result in every single prompt-injection cell, especially under Star and Random topologies. However, it achieves the strongest macro-average performance in every topology. This indicates that \ours provides a more balanced security--utility trade-off across heterogeneous attack types, rather than overfitting to a single attack scenario. In contrast, reactive defenses such as G-Safeguard, BlindGuard, and XG-Guard can reduce ASR in some cases, but their average ACC is consistently lower, suggesting that agent isolation or link pruning may disrupt useful collaboration. By intervening at the message level, \ours better preserves MAS functionality while preventing attack propagation.

\subsection{Detailed robustness analysis across topologies and backbone LLMs}
\label{app:asr}

Figure~\ref{fig:asr_app} provides the detailed results corresponding to the robustness and generalization analysis in the main text. We evaluate \ours under four representative communication topologies, including Chain, Tree, Star, and Random, and compare it with No Defense, G-Safeguard, BlindGuard, and XG-Guard using attack success rate (ASR, lower is better) as the evaluation metric. Figure~\ref{fig:asr_app}(a) reports the ASR under different attack scenarios, including Prompt Injection, Tool Attack, Memory Attack, and Communication Hijacking. The results show that \ours consistently achieves the lowest or near-lowest ASR across different interaction structures, indicating that its protection is not tied to a specific communication topology.

Figure~\ref{fig:asr_app}(b) further examines the generalization ability of \ours across different backbone LLMs, including GPT-4o-mini, DeepSeek-V3.2, Qwen3-30B-A3B, and DeepSeek-V3. Across all backbone models and topologies, \ours maintains a consistently low ASR compared with baseline defenses. These detailed results support the conclusion in the main text that \ours is robust to topology variations and can generalize across heterogeneous LLM backbones.

Figure~\ref{fig:multi-turn_app} presents the detailed results of \ours under multi-turn dialogue settings across four communication topologies, including Chain, Tree, Star, and Random. We evaluate both utility and safety using ACC ($\uparrow$) and ASR ($\downarrow$), respectively, as the number of dialogue turns increases. The results show that \ours maintains consistently high ACC while keeping ASR at a very low level across all topologies. In contrast, baseline methods often suffer from either degraded task performance or increased attack success as the dialogue becomes longer.

Specifically, under Chain, Tree, Star, and Random topologies, \ours achieves final ACC scores of 94.00, 95.00, 95.50, and 92.50, respectively, while keeping the corresponding ASR at only 3.13, 3.44, 1.88, and 5.00. These results indicate that \ours remains effective in agent interactions, where malicious instructions or unsafe behaviors may accumulate across multiple dialogue turns. The consistently low ASR and stable ACC further demonstrate that \ours provides robust protection without compromising the agents' task-solving ability.

\section{Use of LLMs}
LLMs were employed in this work to assist in writing and polishing the paper. All substantive research content, results, and analyses were independently conducted by the authors. The use of LLMs was limited to improving clarity, grammar, and presentation, and did not influence any scientific findings. Detailed experimental methods and results are fully described in the main text of the paper.

% \section{Example Appendix}

\end{document}